\documentclass[10pt,journal,epsfig,twoside]{IEEEtran}
\usepackage{graphicx}
\usepackage{subfigure,color}
\usepackage{amssymb,epstopdf}
\usepackage{cite,comment}
\usepackage{amsmath}
\usepackage{bm,comment}
\usepackage{algorithm}
\usepackage{algpseudocode}
\makeatletter

\newcommand{\Rmnum}[1]{\expandafter\@slowromancap\romannumeral #1@}
\newcommand{\mv}[1]{\mbox{\boldmath{$ #1 $}}}

\makeatother

\newtheorem{proposition}{Proposition}
\newtheorem{remark}{\underline{Remark}}

\begin{document}
\title{Performance Analysis and User Association Optimization for Wireless Network Aided by Multiple Intelligent Reflecting Surfaces}
\author{Weidong Mei and Rui Zhang, \IEEEmembership{Fellow, IEEE}\vspace{-12pt}
\thanks{This work has been presented in part at the IEEE Global Communications Conference, December 7-11, 2020, Taipei, Taiwan\cite{joint2020mei}.}
\thanks{W. Mei is with the NUS Graduate School, National University of Singapore, Singapore 119077, and also with the Department of Electrical and Computer Engineering, National University of Singapore, Singapore 117583 (e-mail: wmei@u.nus.edu).}
\thanks{R. Zhang is with the Department of Electrical and Computer Engineering, National University of Singapore, Singapore 117583 (e-mail: elezhang@nus.edu.sg).}}
\markboth{IEEE TRANSACTIONS ON COMMUNICATIONS}{}
\maketitle

\begin{abstract}
Intelligent reflecting surface (IRS) is deemed as a promising solution to improve the spectral and energy efficiency of wireless communications cost-effectively. In this paper, we consider a wireless network where multiple base stations (BSs) serve their respective users with the aid of distributed IRSs in the downlink communication. Specifically, each IRS assists in the transmission from its associated BS to user via passive beamforming, while in the meantime, it also randomly scatters the signals from other co-channel BSs, thus resulting in additional signal as well as interference paths in the network. As such, a new IRS-user/BS association problem arises pertaining to optimally balance the passive beamforming gains from all IRSs among different BS-user communication links. To address this new problem, we first derive a tractable lower bound of the average signal-to-interference-plus-noise ratio (SINR) at the receiver of each user, termed average-signal-to-average-interference-plus-noise ratio (ASAINR), based on which two ASAINR balancing problems are formulated to maximize the minimum ASAINR among all users by optimizing the IRS-user associations without and with BS transmit power control, respectively. We also characterize the scaling behavior of user ASAINRs with the increasing number of IRS reflecting elements to investigate the different effects of IRS-reflected signal versus interference power. Moreover, to solve the two ASAINR balancing problems that are both non-convex optimization problems, we propose an optimal solution to the problem without BS power control and low-complexity suboptimal solutions to both problems by applying the branch-and-bound method and exploiting new properties of the IRS-user associations, respectively. Numerical results verify our performance analysis and also demonstrate significant performance gains of the proposed solutions over benchmark schemes.
\end{abstract}
\begin{IEEEkeywords}
Intelligent reflecting surface (IRS), passive beamforming, random scattering, IRS association, SINR balancing, power control.
\end{IEEEkeywords}

\section{Introduction}
The use of explosively growing active nodes such as base station (BS), relay and centralized/distributed antennas in today's wireless network has incurred increasingly more energy consumption and higher hardware cost. In view of this issue, both academia and industry have been exploring new and more sustainable solutions to enhance wireless network performance yet at affordable cost. Recently, intelligent reflecting surface (IRS) or its various equivalents (e.g., reconfigurable intelligent surface) has emerged as an appealing candidate thanks to its promising passive beamforming gains brought to wireless communications\cite{wu2020intelligent,wu2019towards,basar2019wireless} as well as lower hardware cost and energy consumption as compared to conventional phased arrays\cite{dai2020reconfigurable}. IRS is typically a planar surface comprising a massive number of sub-wavelength sized, passive, and tunable reflecting elements. By adjusting the reflection amplitude/phase-shift of individual elements, they can jointly alter the strength/direction of the reflected signal by IRS for achieving various purposes, such as beamforming, interference nulling, spatial multiplexing, etc. For example, the signal reflected by an IRS can be constructively/destructively combined with those propagating through other paths at an intended/unintended receiver to enhance/suppress its received signal power.

\begin{figure}[!t]
\centering
\includegraphics[width=3.5in]{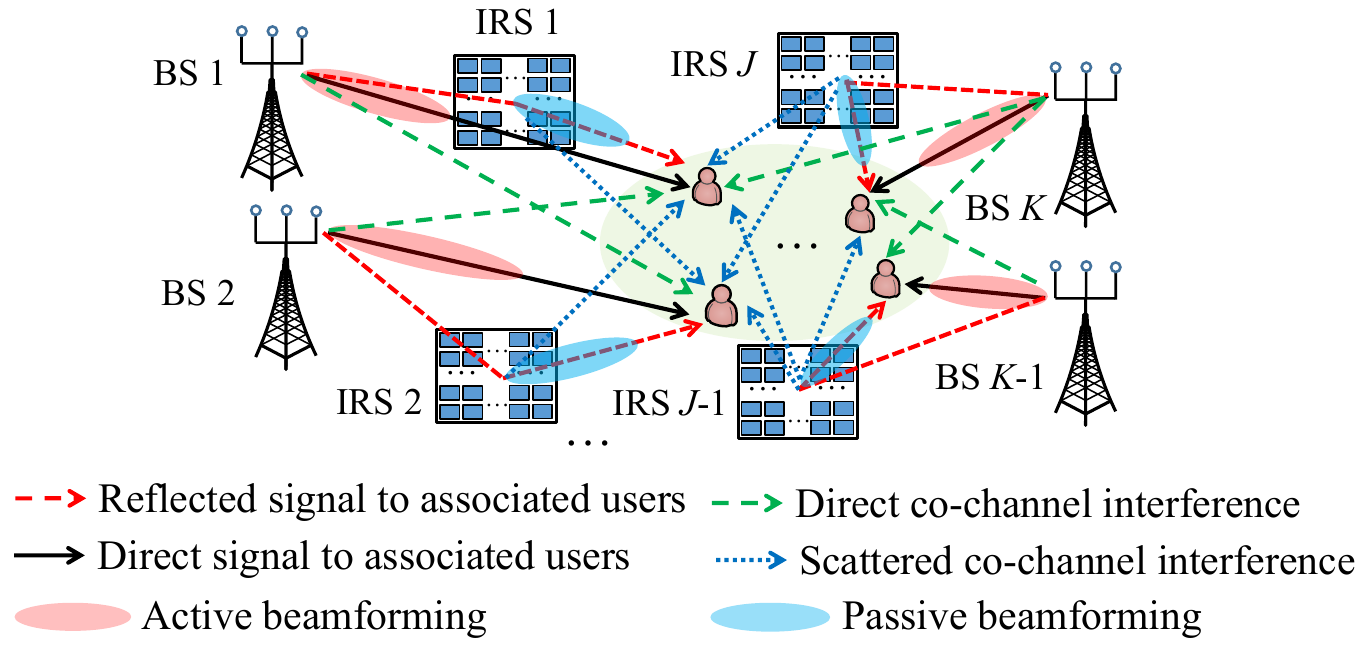}
\DeclareGraphicsExtensions.\vspace{-9pt}
\caption{A multi-IRS aided multi-antenna wireless network.}\label{IRS_Assoc}
\vspace{-9pt}
\end{figure}

The promising and multifarious benefits of IRS have spurred great enthusiasm in investigating optimal IRS reflection or passive/reflect beamforming designs in various IRS-aided wireless systems\cite{wu2019intelligent,zhang2019capacity,mu2020exploiting,wu2020weighted,huang2019reconfigurable,di2019hybrid,yan2019passive,kammoun2020asymptotic,ozdogan2020using,lu2021aerial,kishk2020exploiting,di2019reflection}. It has been shown that passive beamforming can dramatically improve the wireless system performance as compared to the traditional system without IRS. However, most of the existing works on IRS have focused on the {\it link-level} performance optimization, while there is few work on the performance optimization in the general multi-IRS aided wireless network. In this paper, we aim to investigate the {\it network-level} performance optimization for a general wireless network consisting of multiple BSs serving their respective users aided by distributed IRSs, as shown in Fig.\,\ref{IRS_Assoc}. In particular, each IRS adapts its passive beamforming to assist in the downlink communication from its associated BS to user while randomly scattering the signals from other non-associated BSs, which thus results in additional signal as well as interference paths in the network. As such, a new IRS-user/BS association problem arises in our considered multi-IRS aided wireless network. Specifically, with any given BS transmit precoding, assigning more IRSs to one user helps improve its signal-to-interference-plus-noise ratio (SINR) due to the higher passive beamforming gain for enhancing its received signal power. However, this will compromise the SINRs of other users as the total number of distributed IRSs in the network is fixed. In addition, the IRS-user associations may also impact the signal/interference powers due to the random scattering of non-associated IRSs at each user's receiver, thus affecting its SINR performance. This issue becomes more severe for cell-edge users that connect with different BSs due to their close distances with the same set of nearby IRSs. Furthermore, the IRS-user association design is also coupled with BS transmit precoding, since different BS transmit beamforming may change the optimal IRS-user associations and vice versa. Thus, the IRS-user association design for a multi-IRS aided wireless network is a new and non-trivial problem, which, however, has not been studied in the literature to the authors' best knowledge. It is worth noting that multi-IRS aided wireless network has been recently considered in a handful of related works\cite{pan2020multicell,xie2021max,zhang2020capacity,huang2020decentralized,hua2020intelligent,ni2020resource,lyu2021hybrid}. However, these works assumed given IRS-user/BS associations and thus did not investigate their optimal design along with other key system parameters such as BS transmit precoding or power control.

Motivated by the above, in this paper, we study the IRS-user association design for the downlink communication in a multi-IRS aided multi-antenna wireless network, as shown in Fig.\,\ref{IRS_Assoc}. Each BS serves its associated user by applying the maximum ratio transmission (MRT) precoding. Our main contributions are summarized as follows.
\begin{itemize}
\item First, a tractable lower bound of the average received SINR of each user, termed average-signal-to-average-interference-plus-noise ratio (ASAINR), is derived in closed-form by taking into account the effects of MRT by its associated BS, passive beamforming by associated IRSs, and random scattering by non-associated IRSs. Accordingly, we formulate two ASAINR balancing problems to maximize the minimum ASAINR among all users in the network (referred to as the network common ASAINR) by optimizing the IRS-user associations without and with BS power control, respectively. Performance analysis is also provided to characterize the scaling behavior of user ASAINRs with respect to (w.r.t.) the increasing number of reflecting elements per IRS, denoted by $M$, to draw useful insights. It is shown that when at least one IRS is associated with each user to exploit the passive beamforming gain, its ASAINR linearly increases with $M$ even in the interference-limited regime. However, if there is no IRS associated with the user (i.e., its received signals from all IRSs are randomly scattered), its ASAINR is severely limited by the multi-user interference as in the conventional network without IRS, regardless of $M$. Based on this result, we further characterize the conditions on $M$ under which the maximum network common ASAINR in the considered multi-IRS aided network is ensured to be higher than that in the conventional wireless network without IRS, for both ASAINR balancing problems with or without BS power control.

\item Furthermore, as the formulated ASAINR balancing problems are non-convex mixed-integer nonlinear programming (MINLP), they are difficult to solve in general. Although a full enumeration over all IRS-user associations is able to solve them optimally, the required complexity is prohibitive in practice with a large number of IRSs/users. Moreover, the commonly used alternating optimization (AO) approach for alternately solving one of the BS power control and IRS-user associations with the other being fixed, is shown to be ineffective for our formulated problems due to an early-termination issue. To tackle the above difficulties, we propose an optimal solution for the case without BS power control by reformulating the problem into an equivalent mixed-integer linear programming (MILP) and solving it via the branch-and-bound (BB) method. In addition, suboptimal but low-complexity solutions to both problems are proposed by iteratively switching the IRS association among users, which can overcome the early-termination issue with AO. Last, numerical results verify our performance analysis and also demonstrate that the two proposed suboptimal solutions perform nearly optimal and also significantly outperform some heuristic schemes. Moreover, it is revealed that the IRS-user associations in general have different effects on the achievable network common ASAINR with versus without BS power control.
\end{itemize}

The rest of this paper is organized as follows. Section \Rmnum{2} presents the system model and problem formulation. Section \Rmnum{3} presents analytical results on the user ASAINRs and the maximum network common ASAINR. Section \Rmnum{4} and Section \Rmnum{5} present our proposed solutions to the two ASAINR balancing problems without and with BS power control, respectively. Section \Rmnum{6} presents numerical results to evaluate the performance of the proposed algorithms. Finally, Section \Rmnum{7} concludes the paper and discusses future work.

{\it Notations:} Bold symbols in capital letter and small letter denote matrices and vectors, respectively. The vectorization, transpose and conjugate transpose of a matrix are denoted as ${\rm{vec}}(\cdot)$, ${(\cdot)}^{T}$ and ${(\cdot)}^{H}$, respectively. $A_{i,j}$ denotes the entry of a matrix $\mv A$ in the $i$-th row and the $j$-th column, while $a_i$ denotes the $i$-th entry of a vector $\mv a$. ${\mathbb{R}}^n$ (${\mathbb{C}}^n$) denotes the set of real (complex) vectors of length $n$. For a complex number $s$, $s^*$, $\lvert s \rvert$ and $\angle s$ denote its conjugate, amplitude and phase, respectively, and $s \sim \mathcal{CN}(\mu,\sigma^2)$ means that it is a circularly symmetric complex Gaussian (CSCG) random variable with mean $\mu$ and variance $\sigma^2$. For a vector ${\mv a} \in {\mathbb{C}}^n$, ${\rm diag}({\mv a})$ denotes an $n \times n$ diagonal matrix whose entries are the elements of $\mv a$. ${\mathbb E}[\cdot]$ denotes the expected value of random variables. $\lfloor \cdot \rfloor$ denotes the greatest integer less than or equal to its argument. ${\mv e}_k \in {\mathbb{R}}^n$ denotes the $k$-th column of the identity matrix ${\mv I}_n$. $\emptyset$ denotes an empty set. $\lvert A \rvert$ denotes the cardinality of a set $A$. $j$ denotes the imaginary unit, i.e., $j^2=-1$. For two sets $A$ and $B$, $A \backslash B$ denotes the set of elements that belong to $A$ but are not in $B$. ${\cal O}(\cdot)$ denotes the Landau's symbol to describe the order of convergence as well as complexity. $\Gamma(\cdot)$ denotes the gamma function.

\section{System Model and Problem Formulation}
\subsection{System Model}
As shown in Fig.\,\ref{IRS_Assoc}, we consider the downlink communication in a wireless network, where $K$ multi-antenna BSs serve a cluster of single-antenna users at their cell edge with the help of $J$ IRSs. Each BS and IRS are equipped with $L$ antennas and $M$ passive reflecting elements, respectively. Note that this work can be similarly extended to the uplink communication as well. As the users in the cluster may be close to each other in locations, to avoid the strong intra-cell interference due to their spatially correlated channels with their associated BS, each BS is assumed to serve only one user in the cluster at one time. As such, in this paper, we focus on the $K$ cell-edge users each served by one of the $K$ BSs and denote the sets of BSs/users, IRSs and reflecting elements per IRS as ${\cal K}\triangleq \{1,2,\cdots,K\}$, ${\cal J}\triangleq \{1,2,\cdots,j\}$ and ${\cal M}\triangleq \{1,2,\cdots,M\}$, respectively. Let $\theta_{j,m} \in [0, 2\pi]$ denote the phase shift of the $m$-th element of IRS $j, j \in {\cal J}$. Then its diagonal reflection matrix is expressed as ${\mv \Phi}_j={\rm diag}\{e^{j\theta_{j,1}},e^{j\theta_{j,2}},\cdots,e^{j\theta_{j,M}}\}$, where we have assumed full signal reflection at each IRS element for achieving maximum reflected signal power as well as ease of hardware implementation. Moreover, we only consider the signal paths from any BS to any user that are reflected by at most one IRS, since after multiple IRS reflections, the reflected signal will be substantially attenuated due to the multiplicative path loss of the reflected link\cite{wu2019towards}. For the purpose of exposition, we consider a narrow-band system with frequency-flat channels, while the results can be extended to the more general broadband system with frequency-selective channels via orthogonal frequency-division multiple access (OFDMA)-based joint resource allocation\cite{yang2020irs}.

Let ${\mv f}^H_{n,k}=[f_{n,k,1},f_{n,k,2},\cdots,f_{n,k,L}] \in {\mathbb C}^{1 \times L}, n,k \in {\cal K}$ denote the (direct) channel from BS $n$ to user $k$ (i.e, without the reflection by any IRS), ${\mv H}_{n,j}=[{\mv h}_{n,j,1},{\mv h}_{n,j,2},\cdots,{\mv h}_{n,j,M}]^H \in {\mathbb C}^{M \times L}, n \in {\cal K}, j \in {\cal J}$ denote that from BS $n$ to IRS $j$, and ${\mv g}_{j,k}^{H}=[g_{j,k,1},g_{j,k,2},\cdots,g_{j,k,M}] \in {\mathbb C}^{1 \times M}, k \in {\cal K}, j \in {\cal J}$ denote that from IRS $j$ to user $k$. In this paper, to obtain tractable performance analysis, we consider a rich-scattering propagation environment in the network, and hence assume independent Rayleigh (small-scale) fading\footnote{It should be mentioned that in practice, independent Rayleigh fading may not be achieved for the elements on IRS due to the limited inter-element spacing\cite{bjornson2020rayleigh}. However, according to our simulation results in Section \ref{sim}, this assumption achieves a high approximation accuracy if the IRS element spacing is sufficiently large.} for all the channels involved with the considered $K$ users, i.e., ${\mv f}_{n,k} \sim {\cal {CN}}({\bf 0},\alpha^2_{n,k}{\mv I}_L)$, ${\rm vec}({\mv H}_{n,j}) \sim {\cal {CN}}({\bf 0},\beta^2_{n,j}{\mv I}_{ML})$, and ${\mv g}_{j,k} \sim {\cal {CN}}({\bf 0},\eta^2_{j,k}{\mv I}_M), n,k \in {\cal K}, j \in {\cal J}$, where $\alpha^2_{n,k}$, $\beta^2_{n,j}$ and $\eta^2_{j,k}$ are the corresponding distance-dependent average power gains. We consider that the BS-user associations have been established based on some practical association rules, e.g., the maximum reference signal received power (RSRP) in Long Term Evolution (LTE)\cite{3GPP38901}. All IRSs can be regarded as random scatterers in the network in this phase, or can be simply switched off. Without loss of generality, we assume that user $k, k \in \cal K$ is associated with BS $k$ and served by its active precoding vector ${\mv w}_{k} \in {\mathbb C}^{L \times 1}$. For analytical tractability and low-complexity design, it is also assumed that BS $k$ applies the MRT precoding based on its direct channel with user $k$, i.e.,
\begin{equation}\label{activeBF}
{\mv w}_k= \frac{{\mv f}_{k,k}}{\lVert {\mv f}_{k,k} \rVert}, k \in {\cal K}.
\end{equation}
Hence, each user can simultaneously reap the active and passive beamforming gains provided by its associated BS and IRSs over the direct and reflected links with them, respectively, as depicted in Fig.\,\ref{IRS_Assoc}. Moreover, as will be shown in Sections \ref{metric} and \ref{sim}, the MRT precoding in (\ref{activeBF}) can yield a tight and tractable lower bound on the ASAINR of user $k$ in our considered scenario under the independent Rayleigh fading channel assumption. Based on (\ref{activeBF}), for notational simplicity, we can equivalently consider $K$ single-antenna BSs in $\cal K$. In particular, the equivalent direct channel from BS $n$ to user $k$ is expressed as $\tilde f_{n,k}={\mv f}^H_{n,k}{\mv w}_n \in {\mathbb C}, n,k \in {\cal K}$, while that from BS $n$ to IRS $j$ is expressed as $\tilde{\mv h}_{n,j}={\mv H}_{n,j}{\mv w}_n=[\tilde h^*_{n,j,1},\tilde h^*_{n,j,2},\cdots,\tilde h^*_{n,j,M}]^H \in {\mathbb C}^{M \times 1}, n \in {\cal K}, j \in {\cal J}$ with $\tilde h_{n,j,m}={\mv h}^H_{n,j,m}{\mv w}_n, m \in \cal M$. Then, we have $\tilde f_{k,k}={\mv f}^H_{k,k}{\mv w}_k=\lVert {\mv f}_{k,k} \rVert, k \in {\cal K}$; thus, ${\mathbb E}[\lvert\tilde f_{k,k}\rvert^2]={\mathbb E}[\lVert {\mv f}_{k,k} \rVert^2]=L\alpha_{k,k}^2$ and ${\mathbb E}[\tilde f_{k,k}]={\mathbb E}[\lVert {\mv f}_{k,k} \rVert]=\frac{\Gamma(L+\frac{1}{2})}{\Gamma(L)}\alpha_{k,k}$\cite{park1961moments}. Moreover, conditioning on any given ${\mv w}_n$, it can be shown that $\tilde f_{n,k} \sim {\cal {CN}}(0,\alpha^2_{n,k} \lVert {\mv w}_n\rVert^2), n \ne k, n,k \in \cal K$ and $\tilde {\mv h}_{n,j} \sim {\cal {CN}}({\bf 0},\beta^2_{n,j}\lVert {\mv w}_n\rVert^2{\mv I}_M), n \in {\cal K}, j \in {\cal J}$. As $\lVert {\mv w}_n\rVert=1, n \in \cal K$, it follows that ${\mathbb E}[\lvert\tilde f_{n,k}\rvert^2]={\mathbb E}[\alpha^2_{n,k}\lVert {\mv w}_n\rVert^2]=\alpha^2_{n,k}, n \ne k, n,k \in \cal K$ and ${\mathbb E}[\tilde {\mv h}_{n,j}\tilde {\mv h}^H_{n,j}]={\mathbb E}[\beta^2_{n,j}\lVert {\mv w}_n\rVert^2]{\mv I}_M=\beta^2_{n,j}{\mv I}_M, n \in {\cal K}, j \in {\cal J}$.

To assist in the downlink communications between the BSs and their respectively served users, each IRS $j, j \in {\cal J}$ can be associated with one user $k, k \in \cal K$ and adjust its reflection phases accordingly, such that its reflected signal can be constructively combined with the directly transmitted signal from the serving BS of user $k$ (i.e., BS $k$) at the user receiver. Thus, we define the binary variables $\lambda_{j,k}, j \in {\cal J}, k \in {\cal K}$, which indicate that IRS $j$ is associated with user $k$ if $\lambda_{j,k} = 1$; otherwise, $\lambda_{j,k} = 0$. To simplify the IRS phase-shift design and association in practice as well as reduce the channel estimation complexity, we assume that each IRS $j$ can only be associated with at most one user in ${\cal K}$. Thus, we have
\begin{equation}
\sum\limits_{k \in {\cal K}}{\lambda_{j,k}} \le 1, \forall j \in {\cal J}.
\end{equation}

Based on the above, if IRS $j,j \in \cal J$ assists in the downlink communication between BS $k$ and user $k$, i.e., $\lambda_{j,k}=1$, the cascaded BS-IRS-user channel ${\mv g}^H_{j,k}{\mv \Phi}_j\tilde{\mv h}_{k,j}$ should be aligned in phase with the direct BS-user channel $\tilde f_{k,k}$. To this end, the $m$-th phase shift of IRS $j$ should be set as\cite{wu2019intelligent}
\begin{align}
\theta_{j,m}&=\angle{\tilde f_{k,k}}-\angle{g_{j,k,m}}-\angle{\tilde h_{k,j,m}}\nonumber\\
&=-\angle{g_{j,k,m}}-\angle{({\mv h}^H_{k,j,m}{\mv f}_{k,k})}, m \in {\cal M}.\label{ps}
\end{align}
Note that each BS $k$ only needs to collect local channel state information (CSI) on its served users and associated IRSs, i.e., $f_{k,k}$, ${\mv g}_{j,k}$, and ${\mv H}_{k,j}$ with $\lambda_{j,k}=1$, to determine the phase shifts of IRS $j$ in (\ref{ps}), which can be obtained by customized channel estimation schemes for IRS as proposed in \cite{yang2020intelligent,zheng2019intelligent,you2019progressive}. It is also worth noting that the BSs can further optimize the phase shifts of their respective associated IRSs to maximize other utilities for the users, such as their instantaneous SINRs. However, to this end, each BS needs to collect the CSI on other non-associated users and IRSs, which incurs larger overhead and delay. In addition, the corresponding optimal IRS phase shifts are more challenging to obtain and analyze as compared to (\ref{ps}).

Given the phase shifts of all IRSs, they can reflect its intended signals while scattering its unintended signals randomly at the same time. As a result, each user $k, k \in \cal K$, can receive its desired signal via the reflection of all its associated IRSs in $\cal J$, as well as the scattering of all its non-associated IRSs in $\cal J$. By ignoring the signals reflected or scattered by IRSs two or more times, the desired signal received by user $k \in {\cal K}$ from BS $k$ is given by
\begin{align}\label{des0}
y_k&=\tilde f_{k,k}x_k + \Big(\sum\limits_{j \in \cal J}{{\mv g}_{j,k}^{H}{{\mv \Phi}_j}\tilde{\mv h}_{k,j}}\Big)x_k \nonumber\\
&=\tilde f_{k,k}x_k + \Big(\sum\limits_{j \in \cal J}{\lambda_{j,k}{\mv g}_{j,k}^{H}{{\mv \Phi}_j}\tilde{\mv h}_{k,j}}\Big)x_k\nonumber\\
&\qquad\qquad+\Big(\sum\limits_{j \in \cal J}{(1-\lambda_{j,k}){\mv g}_{j,k}^{H}{{\mv \Phi}_j}\tilde{\mv h}_{k,j}}\Big)x_k,
\end{align}
where $x_k$ denotes the transmitted information symbol of user $k$ and satisfies ${\mathbb E}\{\lvert x_k \rvert^2\}=P_k$ with $P_k \ge 0$ being the transmit power of BS $k$. Obviously, for each IRS $j, j \in \cal J$, if it is associated with user $k$, i.e., $\lambda_{j,k}=1$, it only contributes to the second term in (\ref{des0}). Otherwise, it only contributes to the third term in (\ref{des0}). Then, by substituting (\ref{ps}) into the second term in (\ref{des0}), we have
\begin{equation}\label{des}
\small y_k\!=\!\tilde f_{k,k}x_k +\! \Big(\!\sum\limits_{j \in \cal J}{\lambda_{j,k}}Q_{j,k}\Big)x_k\!+\Big(\!\sum\limits_{j \in \cal J}{(1\!-\!\lambda_{j,k}){\mv g}_{j,k}^{H}{{\mv \Phi}_j}\tilde{\mv h}_{k,j}}\!\Big)x_k,\normalsize
\end{equation}
where $Q_{j,k}=\sum\nolimits_{m \in \cal M} {\lvert \tilde h_{k,j,m} \rvert \lvert g_{j,k,m} \rvert}$. 

As noted from (\ref{des}), the information signals reflected by the associated IRSs in $\cal J$ (i.e., the second term in (\ref{des})) are in-phase with that propagated through the direct BS-user link, i.e., $\tilde f_{k,k}x_k=\lVert {\mv f}_{k,k} \rVert x_k$. However, the signals scattered by the non-associated IRSs (i.e., the third term in (\ref{des})) can be combined either constructively or destructively with $\tilde f_{k,k}x_k$. Notice that the reflection matrix of a non-associated IRS $j \in \cal J$ (i.e., with $\lambda_{j,k} =0$) is determined to align another different BS-user direct channel in phase with its corresponding BS-IRS-user cascaded channel. Due to independent $\tilde f_{n,k}$'s, $g_{j,k,m}$'s and $\tilde h_{n,j,m}$'s over $n,k \in \cal K$ and $j \in {\cal J}$, each $\theta_{j,m}$ in ${\mv \Phi}_j$ for a non-associated IRS $j$ can be regarded as a uniformly distributed random variable within $[0,2\pi]$ at user $k$. Under this assumption, to derive the statistics of ${\mv g}_{j,k}^{H}{{\mv \Phi}_j}\tilde{\mv h}_{k,j}$ in (\ref{des}), we define $\hat{\mv h}_{k,j}={{\mv \Phi}_j}\tilde{\mv h}_{k,j}$. Since each $\theta_{j,m}$ in ${\mv \Phi}_j$ is uniformly distributed within $[0,2\pi]$, it follows that $\hat{\mv h}_{k,j} \sim {\cal {CN}}({\bf 0},\beta_{k,j}^2{\mv I}_M)$\cite{wu2019intelligent}. Then, as $\hat{\mv h}_{k,j}$ is independent of ${\mv g}_{j,k}^{H}$, it can be verified that ${\mathbb E}[{\mv g}_{j,k}^{H}{{\mv \Phi}_j}\tilde{\mv h}_{k,j}]={\mathbb E}[{\mv g}_{j,k}^{H}\hat{\mv h}_{k,j}]=0$ and ${\mathbb E}[\lvert{\mv g}_{j,k}^{H}{{\mv \Phi}_j}\tilde{\mv h}_{k,j}\rvert^2]={\mathbb E}[\lvert{\mv g}_{j,k}^{H}\hat{\mv h}_{k,j}\rvert^2]=M\beta_{k,j}^2\eta_{j,k}^2$.

Meanwhile, the other $K-1$ BSs may impose co-channel interference to user $k$ through both their direct channels and cascaded channels via all the IRSs in $\cal J$, regardless of their associated users. The co-channel interference received by user $k$ is given by
\begin{equation}\label{inf}
I_k=\sum\limits_{n \in {\cal K}, n \ne k}\tilde f_{n,k}x_n+\sum\limits_{n \in {\cal K}, n \ne k}\Big(\sum\limits_{j \in \cal J}{\mv g}_{j,k}^{H}{{\mv \Phi}_j}\tilde{\mv h}_{n,j} \Big)x_n.
\end{equation}
Similar to (\ref{des}), since each ${\mv \Phi}_j, j \in \cal J$ is determined independently of either ${\mv g}_{j,k}^{H}$, $\tilde{\mv h}_{n,j}$ or both, the IRS-induced co-channel interference in (\ref{inf}) satisfies ${\mathbb E}[{\mv g}_{j,k}^{H}{{\mv \Phi}_j}\tilde{\mv h}_{n,j}]=0$ and ${\mathbb E}[\lvert {\mv g}_{j,k}^{H}{{\mv \Phi}_j}\tilde{\mv h}_{n,j} \rvert^2]=M\beta_{n,j}^2\eta_{j,k}^2, n \in {\cal K}, n \ne k$.

\subsection{Performance Metric}\label{metric}
Given the above characterization of the IRS-reflected information signal and co-channel interference, we aim to investigate the average user SINR performance in the network. The average receive SINR at the receiver of user $k$ is given by
\begin{equation}\label{sinr}
\tilde\gamma_k={\mathbb E}\left[\frac{\lvert y_k \rvert^2}{\sigma^2+\lvert I_k\rvert^2}\right],
\end{equation}
where $\sigma^2$ is the background noise power, and the expectation is taken over all Rayleigh-fading channels involved. However, it is generally difficult to express the average SINR in (\ref{sinr}) in a tractable form. Therefore, we approximate the average SINR by its lower bound given below,
\begin{equation}\label{sinr.lw}
\tilde\gamma_k={\mathbb E}\left[\lvert y_k \rvert^2 \right]{\mathbb E}\left[\frac{1}{\sigma^2+\lvert I_k\rvert^2}\right] \ge \frac{{\mathbb E}\left[\lvert y_k \rvert^2 \right]}{\sigma^2+{\mathbb E}\left[\lvert I_k\rvert^2\right]} \triangleq \gamma_k,
\end{equation}
where the equality is due to the fact that $y_k$ and $I_k$ are independent, and the inequality holds due to the Jensen's inequality since the function $\frac{1}{x}$ is convex in $x$ for $x > 0$. For convenience, we refer to $\gamma_k$ as the ASAINR of user $k$ in the sequel of this paper. Note that $\gamma_k$ is only related to the large-scale average power gains or statistical CSI, which can be used to determine the IRS-user associations. After the BS-user and IRS-user associations are established, local CSI can be estimated in real time by each BS to determine its active precoding, as well as the phase shifts of its associated IRSs. In particular, the active precoding ${\mv w}_k$ can then be implemented as the MRT based on the effective BS $k$-user $k$ channel, i.e., ${\mv f}^H_{k,k}+\sum\nolimits_{j \in \cal J}{{\mv g}_{j,k}^{H}{{\mv \Phi}_j}{\mv H}_{k,j}}$, instead of using that based on the direct BS $k$-user $k$ channel ${\mv f}^H_{k,k}$ only for the IRS-user association optimization. This will help improve the desired signal power of user $k$ yet without increasing the average interference power from BS $k$ due to its independent channels with different users as well as IRSs. As such, user $k$'s ASAINR with the direct channel based MRT (DC-MRT), $\gamma_k$, offers a tractable lower bound of its ASAINR as well as average SINR with the optimal effective channel based MRT (EC-MRT). It is worth mentioning that similar statistical CSI-based association strategies have been widely applied in practice to determine the BS-user associations, such as the maximum RSRP-based association rule in LTE. Next, we show that each $\gamma_k, k \in \cal K$ can be derived in a closed-form.

First, based on (\ref{des}), it can be shown that
\begin{align}
{\mathbb E}\left[\lvert y_k \rvert^2\right]=&P_k{\mathbb E}\Big[\Big(\tilde f_{k,k} + \sum\limits_{j \in \cal J}\lambda_{j,k}Q_{j,k}\Big)^2\Big]\nonumber\\
&+P_k{\mathbb E}\Big[\Big|\sum\limits_{j \in \cal J}{(1-\lambda_{j,k}){\mv g}_{j,k}^{H}{{\mv \Phi}_j}\tilde{\mv h}_{k,j}}\Big|^2\Big],\label{mean1}
\end{align}
which is due to the independence between the cascaded channels via the associated IRSs and those via the non-associated IRSs. Let $E_1$ and $E_2$ denote the first and the second expectation term in (\ref{mean1}), respectively, i.e., ${\mathbb E}\left[\lvert y_k \rvert^2\right]=P_kE_1+P_kE_2$ and define $q_{n,j,k}=\beta_{n,j}\eta_{j,k}, n,k \in {\cal K}, j \in \cal J$ as the average path gain between BS $n$ and user $k$ via IRS $j$. Then, by expanding $E_1$, we can obtain
\begin{align}
E_1&\!\!=\!{\mathbb E}[\tilde f_{k,k}^2]\!+\!2{\mathbb E}[\tilde f_{k,k}]\Big(\sum\limits_{j \in \cal J}\lambda_{j,k}{\mathbb E}[Q_{j,k}]\Big)\!\!+\!{\mathbb E}\Big[\!\Big(\!\sum\limits_{j \in \cal J}\lambda_{j,k}Q_{j,k}\!\Big)^2\Big]\nonumber\\
&\!\!=\!L\alpha^2_{k,k}+\frac{2\Gamma(L+\frac{1}{2})\alpha_{k,k}}{\Gamma(L)}\Big(\sum\limits_{j \in \cal J}\lambda_{j,k}{\mathbb E}[Q_{j,k}]\Big)\nonumber\\
&\quad+{\mathbb E}\Big[\Big(\sum\limits_{j \in \cal J}\lambda_{j,k}Q_{j,k}\Big)^2\Big],\label{harden0}
\end{align}
where we have exploited ${\mathbb E}[\tilde f_{k,k}^2]=L\alpha_{k,k}^2$ and ${\mathbb E}[\tilde f_{k,k}]=\frac{\Gamma(L+\frac{1}{2})}{\Gamma(L)}\alpha_{k,k}$ given after (\ref{activeBF}). Moreover, it must hold that
\begin{equation}\label{harden1}
{\mathbb E}\left[Q_{j,k}\right]=\sum\limits_{m \in \cal M} {\mathbb E}[\lvert \tilde h_{k,j,m} \rvert] {\mathbb E}[\lvert g_{j,k,m} \rvert]=\frac{M\pi}{4}q_{k,j,k},
\end{equation}
where the first equality is because each $\lvert \tilde h_{k,j,m} \rvert$ and $\lvert g_{j,k,m} \rvert$ are statistically independent, and the second equality is due to the fact that $\lvert g_{j,k,m} \rvert$ follows Rayleigh distribution with mean value of $\frac{\sqrt{\pi}\eta_{j,k}}{2}$. Furthermore, since conditioning on any given ${\mv w}_k$, we have $\tilde h_{k,j,m} \sim {\cal {CN}}(0,\beta^2_{k,j}\lVert {\mv w}_k \rVert^2)$, it follows that ${\mathbb E}[\lvert \tilde h_{k,j,m} \rvert]={\mathbb E}[\frac{\sqrt{\pi}}{2}\beta_{k,j}\lVert {\mv w}_k \rVert]=\frac{\sqrt{\pi}\beta_{k,j}}{2}$. Similarly, by expanding $Q^2_{j,k}$ and exploiting the above facts, we can obtain the second moment of $Q_{j,k}$ as
\begin{equation}\label{harden2}
{\mathbb E}\left[Q^2_{j,k}\right]=\frac{M^2\pi^2}{16}q^2_{k,j,k}+Mq^2_{k,j,k}\left(1-\frac{\pi^2}{16}\right).
\end{equation}
By utilizing (\ref{harden1}) and (\ref{harden2}) in (\ref{harden0}), $E_1$ can be recast as the following closed form,
\begin{align}
	E_1=&L\alpha^2_{k,k}+\frac{\pi\Gamma(L+\frac{1}{2}) M\alpha_{k,k}}{2\Gamma(L)}\sum\limits_{j \in \cal J}\lambda_{j,k}q_{k,j,k}\nonumber\\
	&\!+\!M\left(1\!-\!\frac{\pi^2}{16}\right)\sum\limits_{j \in \cal J}\lambda_{j,k}q^2_{k,j,k}\!+\!\frac{M^2\pi^2}{16}\Big(\sum\limits_{j \in \cal J}\lambda_{j,k}q_{k,j,k}\Big)^2.\nonumber
\end{align}

On the other hand, $E_2$ can be simplified as follows,
\[E_2\!=\!\sum\limits_{j \in \cal J}{(1-\lambda_{j,k}){\mathbb E}[\lvert{\mv g}_{j,k}^{H}{{\mv \Phi}_j}\tilde{\mv h}_{k,j}\rvert^2]}\!=\!M\sum\limits_{j \in \cal J}{(1-\lambda_{j,k})q_{k,j,k}^2},\]
where the first equality is due to the independence among the cascaded channels via different non-associated IRSs and the fact that $(1-\lambda_{j,k})^2=1-\lambda_{j,k}, \forall j,k$, while the second equality is due to ${\mathbb E}[\lvert{\mv g}_{j,k}^{H}{{\mv \Phi}_j}\tilde{\mv h}_{k,j}\rvert^2]=Mq_{k,j,k}^2$. By combining $E_1$ with $E_2$ and rearranging their terms, we can obtain the effective channel power between BS $k$ and user $k$ as
\begin{align}
\tilde\alpha_{k,k}^2(\mv\lambda_k) &\triangleq E_1\!+\!E_2\nonumber\\
&=L\alpha_{k,k}^2\!+\!M\Big(\sum\limits_{j \in \cal J}q_{k,j,k}^2\!+\sum\limits_{j \in \cal J}\lambda_{j,k}A_{j,k} \Big)\nonumber\\
&\quad+\frac{M^2\pi^2}{16}\Big(\sum\limits_{j \in \cal J}\lambda_{j,k}q_{k,j,k}\Big)^2,\label{effch1}
\end{align}
with $\mv\lambda_k \triangleq [\lambda_{j,k}]_{j \in \cal J}, k \in \cal K$ and $A_{j,k}\triangleq \frac{\pi\Gamma(L+\frac{1}{2}) \alpha_{k,k}}{2\Gamma(L)}q_{k,j,k}-\frac{\pi^2}{16}q^2_{k,j,k}, k \in {\cal K}, j \in \cal J$. Notice that (\ref{effch1}) consists of three terms, where the first term linearly increases with the active beamforming gain of $L$ by BS $k$, the second term linearly increases with $M$ and the third term increases quadratically with $M$, which are due to random scattering by non-associated IRSs and passive beamforming by associated IRSs, respectively. If $\sum\nolimits_{j \in \cal J}\lambda_{j,k}>0$, i.e., at least one IRS is associated with user $k$, the effective BS $k$-user $k$ channel power can be significantly improved compared to that without any associated IRS due to the IRS passive beamforming gain offered by the associated IRS, which increases quadratically (versus linearly) with $M$. On the other hand, if $\lambda_{j,k}=0, \forall j \in \cal J$, i.e., there is no IRS associated with user $k$, the effective BS $k$-user $k$ channel power becomes
\begingroup
\allowdisplaybreaks
\begin{equation}\label{effch3}
\tilde\alpha_{k,k}^2=L\alpha_{k,k}^2+M\sum\limits_{j \in \cal J}q_{k,j,k}^2,
\end{equation}
which increases only linearly with $M$.

Similarly, after some mathematical manipulations, we can obtain the average interference power at the receiver of user $k$ as
\begin{align}\label{mean2}
{\mathbb E}\left[\lvert I_k \rvert^2\right]&=\sum\limits_{n \in {\cal K}, n \ne k}P_n\Big(\alpha^2_{n,k}+\sum\limits_{j \in \cal J}Mq_{n,j,k}^2\Big)\nonumber\\
&=\sum\limits_{n \in {\cal K}, n \ne k}P_n\nu^2_{n,k},
\end{align}
where we have exploited the fact that ${\mathbb E}[\lvert \tilde f_{n,k} \rvert^2]=\alpha^2_{n,k}$ and
\begin{equation}\label{effch2}
\nu^2_{n,k} \triangleq \alpha^2_{n,k}+M\sum\limits_{j \in \cal J}q_{n,j,k}^2, n \in {\cal K}, n \ne k
\end{equation}
is the effective channel power between the co-channel BS $n$ and user $k$. It is shown from (\ref{effch2}) that $\nu^2_{n,k}$ increases linearly with $M$ and is regardless of IRS-user associations $\mv\lambda_k$'s, as all IRSs can be treated as random scatterers in reflecting co-channel interference, as shown in (\ref{inf}). It is also worth noting that unlike $\tilde\alpha_{k,k}^2(\mv\lambda_k)$ in (\ref{effch1}), $\nu^2_{n,k}$ is regardless of $L$. This implies that as $L$ becomes practically large (e.g., in the massive multiple-input multiple-output (MIMO) system), $\tilde\alpha_{k,k}^2(\mv\lambda_k)$ can overwhelm $\nu^2_{n,k}$ even without IRS in the network or equivalently, $M=0$.

By substituting (\ref{mean1}) and (\ref{mean2}) into (\ref{sinr.lw}), the ASAINR achievable by user $k$ is expressed as
\begin{equation}\label{mean3}
\gamma_k(\mv\lambda_k)=\frac{P_k\tilde\alpha^2_{k,k}(\mv\lambda_k)}{\sigma^2+\sum\limits_{n \in {\cal K}, n \ne k}P_n\nu^2_{n,k}}, k \in \cal K.
\end{equation}
It is noted that compared to the traditional wireless network without any IRS deployed, the user ASAINRs in (\ref{mean3}) depend on the IRS-user associations $\mv\lambda_k$'s via the effective BS $k$-user $k$ channel power gain, $\tilde\alpha^2_{k,k}(\mv\lambda_k)$, in addition to the BS transmit powers $\{P_k\}$. Furthermore, it follows from (\ref{effch1}) and (\ref{effch2}) that $\gamma_k(\mv\lambda_k)$ always linearly increases with the number of BS antennas $L$, while its scaling behavior with increasing the number of IRS reflecting elements $M$ will be analyzed later in Section \ref{pa} under different scenarios. 
\begin{remark}
The above results are extensible to the practical scenario where each BS $k, k \in \cal K$ also serves other users outside the cell-edge cluster, if we assume that their intended signals are transmitted in the null-space of the effective BS $k$-user $k$ channel by proper precoding at BS $k$. The only difference is that each cell-edge user would receive additional direct and scattered interferences from the co-channel BSs serving their respective non-cell-edge users, while their statistics can be obtained similarly as in (\ref{mean2}).
\end{remark}

\subsection{Problem Formulation}\label{pf}
In this paper, we formulate two ASAINR balancing problems to maximize the minimum receive ASAINR among all users in $\cal K$, i.e., $\mathop{\min}\nolimits_{k \in {\cal K}} \gamma_k(\mv\lambda_k)$, termed as network common ASAINR, by optimizing the IRS-user associations with and without BS power control, respectively.

First, in the case without BS power control, the IRS-user associations, ${\mv \Lambda}\triangleq [{\mv \lambda}_1,{\mv \lambda}_2,\cdots,{\mv \lambda}_K] \in \{0,1\}^{J \times K}$, are optimized with fixed BS transmit powers, ${\mv P} \triangleq [P_1,P_2,\cdots,P_K]^T \in {\mathbb R}^{K \times 1}$. Let $P_{\max}$ denote the maximum transmit power of each BS to serve each user. For convenience, we assume that $P_k=P_{\max}, k \in \cal K$, in this problem. Then, the IRS-user association optimization problem is formulated as
\begin{align}
{\text{(P1)}}\; \gamma^*_{c,1}=\mathop {\max}\limits_{{\mv \Lambda}}&\; \mathop {\min}\limits_{k \in {\cal K}}\; \frac{\tilde\alpha^2_{k,k}({\mv \lambda}_k)}{\zeta_k} \nonumber\\
\text{s.t.}\;\;&\sum\limits_{k \in \cal K}{\lambda_{j,k}} \le 1, \forall j \in {\cal J}, \label{op1a}\\
&\lambda_{j,k} \in \{0,1\}, \forall j \in {\cal J},  k \in {\cal K},\label{op1b}
\end{align}
where $\zeta_k = \frac{1}{P_{\max}}\Big(\sigma^2+P_{\max}\sum\limits_{n \in {\cal K}, n \ne k}\nu^2_{n,k}\Big), k \in \cal K$, are constants and $\gamma^*_{c,1}$ denotes the optimal value of (P1). It is noted that (P1) is a non-convex MINLP problem, which is NP-hard and difficult to be optimally solved. In fact, even by relaxing all integer variables in $\mv\Lambda$ into their continuous counterparts, i.e., $0 \le \lambda_{j,k} \le 1, \forall j,k$ in (P1), (P1) is still a non-convex optimization problem owing to its objective function, where each $\tilde \alpha_{k,k}({\mv \lambda}_k), k \in \cal K$ is convex (instead of concave) in ${\mv \lambda}_k$ (see (\ref{effch1})). Although the optimal solution to (P1) can be derived by enumerating all feasible IRS-user associations, this incurs a worst-case complexity in the order of ${\cal O}\left(K^J\right)$, which is prohibitive if $J$ is practically large. In Section \ref{p1}, we will optimally solve (P1) based on the BB method, which is more efficient than the full enumeration by properly discarding some solution sets that cannot yield the optimal solution to (P1). Moreover, a suboptimal but low-complexity successive refinement algorithm is proposed to solve (P1) more efficiently.

Next, we consider the ASAINR balancing problem when the IRS-user associations ${\mv \Lambda}$ and the BS transmit powers ${\mv P}$ are jointly optimized to maximize the network common ASAINR, i.e.,
\begin{align}
{\text{(P2)}}\; \gamma^*_{c,2}=\mathop {\max}\limits_{{\mv \Lambda},{\mv P}}&\; \mathop {\min}\limits_{k \in {\cal K}}\; \frac{P_k\tilde\alpha^2_{k,k}({\mv \lambda}_k)}{\sigma^2+\sum\limits_{n \in {\cal K}, n \ne k}P_n\nu^2_{n,k}} \nonumber\\
\text{s.t.}\;\;&\sum\limits_{k \in \cal K}{\lambda_{j,k}} \le 1, \forall j \in {\cal J}, \label{op2a}\\
&0 \le P_k \le P_{\max}, \forall k \in {\cal K},\label{op2b}\\
&\lambda_{j,k} \in \{0,1\}, \forall j \in {\cal J},  k \in {\cal K},\label{op2c}
\end{align}
where $\gamma^*_{c,2}$ denotes the optimal value of (P2); evidently, $\gamma^*_{c,2} \ge \gamma^*_{c,1}$.

Compared with (P1), (P2) is also a non-convex MINLP problem but involves coupling variables $\mv\Lambda$ and $\mv P$, which is thus more challenging to be solved. Fortunately, with any fixed IRS-user associations ${\mv\Lambda}$, (P1) reduces to the conventional SINR balancing problem for the benchmark system without IRS, for which the optimal $\mv P$ can be derived in an analytical form\cite{tan2012fast}. To show the optimal $\mv P$, we first define a channel gain ratio matrix $\tilde{\mv F}(\mv\Lambda) \in {\mathbb R}^{K \times K}$ with entries
\begin{equation}
\tilde F_{k,n}({\mv \lambda}_k)=
\begin{cases}
	\frac{\nu^2_{n,k}}{\tilde\alpha^2_{k,k}({\mv \lambda}_k)}, \;&\text{if\;}n \ne k\\
	0, \;&\text{otherwise},
\end{cases}k \in {\cal K}, n \in \cal K,
\end{equation}
and a noise-to-channel-gain ratio vector $\tilde{\mv v}(\mv\Lambda) \in {\mathbb R}^{K \times 1}$ with entries $\tilde v_k({\mv \lambda}_k) = \frac{\sigma^2}{\tilde\alpha_{k,k}^2({\mv \lambda}_k)}, k \in \cal K$. Then, given any $\mv\Lambda$, when the network common ASAINR is maximized, all users in $\cal K$ should achieve the same ASAINR given by\cite{tan2012fast}
\begin{equation}\label{comSINR}
\gamma^*_{c,2}(\mv\Lambda)={\frac{1}{\mathop {\max}\limits_{k \in \cal K}\rho\left(\tilde{\mv F}(\mv\Lambda)+\frac{1}{P_{\max}}\tilde{\mv v}(\mv\Lambda){\mv e}^T_k\right)}},
\end{equation}
where $\rho(\cdot)$ denotes the spectral radius of its argument (also known as Perron-Frobenius eigenvalue if its argument is a non-negative matrix). Furthermore, let \[i=\arg \mathop {\max}_{k \in \cal K}\rho\left(\tilde{\mv F}(\mv\Lambda)+\frac{1}{P_{\max}}\tilde{\mv v}(\mv\Lambda){\mv e}^T_k\right).\]Then, the optimal transmit powers of all BSs in $\cal K$, denoted as ${\mv P}(\mv\Lambda)$, are given by
\begin{equation}\label{pwctrl}
	{\mv P}(\mv\Lambda)=t{\mv x}\left(\tilde{\mv F}(\mv\Lambda)+(1/P_{\max})\tilde{\mv v}(\mv\Lambda){\mv e}^T_i\right),
\end{equation}
where ${\mv x}(\cdot)$ is the Perron left eigenvector of its argument, and $t=P_{\max}/x_i$. As such, BS $i$ should transmit at the maximum power $P_{\max}$, while other BSs should set their transmit powers no greater than $P_{\max}$ in general.

Based on the above, the optimal solution to (P2) can be obtained by enumerating all feasible IRS-user associations, then computing and comparing their corresponding optimal values based on (\ref{comSINR}). Note that (\ref{comSINR}) involves computing the spectral radii of $K$ $K$-by-$K$ matrices, which results in a complexity in the order of ${\cal O}(K^4)$. As such, the overall complexity of this full enumeration is ${\cal O}(K^{J+4})$, which is still prohibitive if $J$ is practically large. To reduce the computational complexity, a straightforward approach is to apply the AO to (P2) by iteratively optimizing each of $\mv\Lambda$ and $\mv P$ with the other being fixed. In particular, for any given $\mv\Lambda$, the optimal $\mv P$ is given by (\ref{pwctrl}); whereas for any given $\mv P$, the optimal $\mv \Lambda$ can be derived by solving (P1) with the BB-based algorithm (to be specified in Section \ref{bnb}). However, we argue that such an AO algorithm is ineffective to solve (P2), as stated in the following proposition.
\begin{proposition}\label{AOineff}
	If the AO algorithm is utilized to solve (P2), it will terminate with the IRS-user associations updated at most once.
\end{proposition}
\begin{IEEEproof}
Please refer to Appendix \ref{appendixA}.
\end{IEEEproof}

Proposition \ref{AOineff} indicates that if the AO algorithm is applied to solve (P1), its performance may not be good as only one iteration is executed. This is due to the fact that $\mv\Lambda$ only affects the information signal power but not interference power for each user. Particularly, if $\mv\Lambda$ or $\mv P$ is not properly initialized, it becomes more prone to getting trapped at undesired suboptimal solutions. For example, if $\mv\Lambda$ is initialized such that all IRSs are assigned to the same single user, then the AO algorithm will not be able to update $\mv\Lambda$ further, since it is impossible to improve all ${\tilde\alpha}^2_{k,k}$'s at the same time by tuning the IRS-user associations. As such, more effective methods than the AO algorithm are needed to solve (P2). In Section \ref{joint}, we propose a sequential update algorithm to solve (P1), which is shown able to avoid the above early-termination issue with the AO.

\section{Performance Analysis}\label{pa}
In this section, to reveal valuable insights into the considered multi-IRS aided wireless network, we perform theoretical analysis to characterize the performance of user ASAINRs in (\ref{mean3}) and the maximum network common ASAINRs (i.e., optimal values of (P1) and (P2)) w.r.t. the number of IRS reflecting elements, $M$. Moreover, we compare their performance with the ASAINR performance achieved by the benchmark system without using IRS.

\subsection{Individual User ASAINR versus $M$}\label{pa1}
First, we characterize the performance of each $\gamma_k(\mv\lambda_k), k \in \cal K$ in (\ref{mean3}) w.r.t. $M$. Notice that for the benchmark system without IRS, the corresponding user ASAINRs can be obtained by setting $M=0$ in (\ref{mean3}) and thus given by
\begin{equation}\label{mean4}
\gamma_{k,0}=\frac{P_kL\alpha^2_{k,k}}{\sigma^2+\sum\limits_{n \in {\cal K}, n \ne k}P_n\alpha^2_{n,k}}, k \in \cal K.
\end{equation}

The following two IRS-user association scenarios are considered. First, if all IRSs in $\cal J$ only randomly scatter the signal from BS $k$ to user $k$ without performing any passive beamforming, i.e., $\lambda_{j,k} =0, \forall j \in \cal J$, the ASAINRs in (\ref{mean3}) become
\begin{equation}\label{limit1}
	\gamma_k = \frac{P_k\Big(L\alpha^2_{k,k}+\sum\limits_{j \in \cal J}{Mq_{k,j,k}^2}\Big)}{\sigma^2+\sum\limits_{n \in {\cal K}, n \ne k}P_n\Big(\alpha^2_{n,k}+\sum\limits_{j \in \cal J}Mq_{n,j,k}^2\Big)}, k \in \cal K.
\end{equation}
By taking the derivative of $\gamma_k$ in (\ref{limit1}) w.r.t. $M$, the following proposition is obtained.
\begin{proposition}\label{wobf}
	Each $\gamma_k$ in (\ref{limit1}) is monotonically increasing with $M$ if
\begin{equation}\label{mono1}
	\frac{\sum\limits_{j \in \cal J}{q_{k,j,k}^2}}{\sum\limits_{n \in {\cal K}, n \ne k}P_n\sum\limits_{j \in \cal J}q_{n,j,k}^2}>\frac{L\alpha^2_{k,k}}{\sigma^2+\sum\limits_{n \in {\cal K}, n \ne k}P_n\alpha^2_{n,k}}, k \in \cal K.
\end{equation}
Otherwise, it is monotonically decreasing with $M$. Moreover, with any given $L$, when $M$ is sufficiently large, i.e., $M \rightarrow \infty$, we have
\begin{equation}\label{mono2}
\gamma_k \rightarrow \frac{P_k\sum\limits_{j \in \cal J}q_{k,j,k}^2}{\sum\limits_{n \in {\cal K}, n \ne k}P_n\sum\limits_{j \in \cal J}q_{n,j,k}^2}, k \in \cal K.
\end{equation}
\end{proposition}

It is noted that the condition in (\ref{mono1}) is regardless of $M$. This indicates that if $\lambda_{j,k} =0, \forall j \in \cal J$ and the conditions in (\ref{mono1}) are not met, increasing $M$ will even yield worse ASAINRs as compared to $\gamma_{k,0}$'s in the benchmark system without using IRS. Moreover, it is noted that each $\gamma_k, k \in \cal K$ is bounded from above in (\ref{mono2}) as $M \rightarrow \infty$. This is expected as in this case, increasing $M$ enhances $\tilde\alpha^2_{k,k}$ and $\nu^2_{n,k}$ linearly at the same time, as previously shown in (\ref{effch3}) and (\ref{effch2}), respectively.

In contrast, in the second scenario, suppose that there is (at least) one IRS in $\cal J$ (say, IRS $i$) that is associated with user $k, k \in \cal K$, i.e., $\lambda_{i,k}=1$ and $\lambda_{j,k}=0, j \ne i, j \in \cal J$. Then, the user ASAINRs in (\ref{mean3}) can be written as
\begin{equation}\label{limit2}
\gamma_k\!=\!\frac{P_k\Big(L\alpha_{k,k}^2\!+\!M\Big(\sum\limits_{j \in \cal J}q_{k,j,k}^2\!+\!A_{i,k}\Big)\!+\!\frac{M^2\pi^2}{16}q_{k,i,k}^2\Big)}{\sigma^2+\sum\limits_{n \in {\cal K}, n \ne k}P_n\Big(\alpha^2_{n,k}+\sum\limits_{j \in \cal J}Mq_{n,j,k}^2\Big)}, k \in \cal K.
\end{equation}
Similarly, by taking the derivative of each $\gamma_k$ in (\ref{limit2}) w.r.t. $M$, we obtain the following proposition.
\begin{proposition}\label{wbf}
	Define 
	\begin{align}
		&B_k=\sum\limits_{j \in \cal J}q_{k,j,k}^2+A_{i,k},\;C_k=\sum\limits_{n \in {\cal K}, n \ne k}P_n\sum\limits_{j \in \cal J}q_{n,j,k}^2,\nonumber\\
		&D_k=\sigma^2+\sum\limits_{n \in {\cal K}, n \ne k}P_n\alpha^2_{n,k}, k \in \cal K,\nonumber
	\end{align}
which correspond to the scattered information signal power, scattered interference power and interference power over the direct links plus noise power received at user $k$, respectively.
	Then, if $C_kL\alpha^2_{k,k} \le B_kD_k$, $\gamma_k$ in (\ref{limit2}) monotonically increases with $M$. Otherwise, it first decreases with $M$ when $M \le \sqrt{\frac{D^2_k}{C^2_k}+\frac{16(C_kL\alpha^2_{k,k}-B_kD_k)}{\pi^2q^2_{k,i,k}C_k}}-\frac{D_k}{C_k}$ and then increases with $M$. Moreover, with any given $L$, as $M$ is sufficiently large, i.e., $M \rightarrow \infty$, we have
\begin{equation}\label{mono4}
	\gamma_k \rightarrow\frac{P_k(\pi^2q_{k,i,k}^2M+16B_k)}{16C_k} = {\cal O}(M), k \in \cal K.
\end{equation}
\end{proposition}

It follows from Proposition \ref{wbf} that even with IRS passive beamforming, increasing $M$ may not lead to a better ASAINR performance if $M$ is small. While as $M$ increases, the user ASAINRs will be improved, as the quadratic growth in the received signal power overwhelms the linear growth in the received interference power for each $\gamma_k, k \in \cal K$. In particular, as $M \rightarrow \infty$, it is observed from (\ref{mono4}) that the user ASAINRs will linearly increase with $M$. This is similar to the effect of $L$ on user ASAINRs (see (\ref{mean3})) but fundamentally different from the case with IRS random scattering only (see (\ref{mono2})), which is limited by the scattered multi-user interference. Nonetheless, it is worth noting that with increasing $M$, the scaling behavior of $\gamma_k$ in (\ref{mono4}) is different from that of the effective channel power $\tilde\alpha_{k,k}^2(\mv\lambda_k)$ in (\ref{effch1}) (linear versus quadratic). This is due to the existence of scattered co-channel interference which linearly increases with $M$, as shown in (\ref{mean2}) and (\ref{effch2}). Moreover, from (\ref{limit2}), it follows that $\gamma_k \ge \gamma_{k.0}$ in the no-IRS benchmark system if $M$ satisfies
\begin{equation}\label{thres1}
	M > \mathop {\max}\left\{\frac{16}{\pi^2}\left(\frac{L\alpha^2_{k,k}C_k}{q_{k,i,k}^2D_k}-\frac{B_k}{q_{k,i,k}^2}\right),0\right\}.
\end{equation}

By removing the term due to IRS random scattering in (\ref{thres1}), we can obtain a simpler relaxed bound on $M$, i.e.,
\begin{equation}\label{thres2}
	M > \frac{16L\alpha^2_{k,k}C_k}{\pi^2q_{k,i,k}^2D_k}=\frac{16L\alpha^2_{k,k}\sum\limits_{n \in {\cal K}, n \ne k}P_n{\sum\limits_{j \in \cal J}q_{n,j,k}^2}}{\pi^2q_{k,i,k}^2\Big(\sigma^2+\sum\limits_{n \in {\cal K}, n \ne k}P_n\alpha^2_{n,k}\Big)}.
\end{equation}
It is observed that with any given BS transmit powers $\mv P$, the bound in (\ref{thres2}) is reduced by decreasing $L$, $\alpha^2_{k,k}$ and $q_{n,j,k}^2$'s and/or increasing $\alpha^2_{n,k}$'s and $q_{k,i,k}^2$. This indicates that for any given $M$, each $\gamma_k$ is more likely to be improved over $\gamma_{k,0}$ due to the passive beamforming by IRS $i$ if the number of BS antennas (or active beamforming gain by each BS) is small, the direct BS $k$-user $k$ link and the reflected BS $n$-user $k$ links are weak (e.g., blocked user in the cell), or the direct BS $n$-user $k$ link and the reflected BS $k$-user $k$ link are strong (e.g, user at the cell edge).

In general, it can be verified that if multiple IRSs in $\cal J$ are associated with user $k$, i.e., $\sum\nolimits_{j \in \cal J}\lambda_{j,k} \ge 1$, the bound in (\ref{thres2}) becomes
\begin{equation}\label{thres2g}
	M > \frac{16L\alpha^2_{k,k}\sum\limits_{n \in {\cal K}, n \ne k}P_n{\sum\limits_{j \in \cal J}q_{n,j,k}^2}}{\pi^2\Big(\sum\limits_{j \in \cal J}\lambda_{j,k}q_{k,j,k}\Big)^2\Big(\sigma^2+\sum\limits_{n \in {\cal K}, n \ne k}P_n\alpha^2_{n,k}\Big)}.
\end{equation}

\textbf{Numerical Example}: To verify our above analysis on the characterization of $\gamma_k$'s under the above two scenarios, we provide the following two numerical examples. In both examples, we set $L=1$, $K=2$ and $J=1$ and focus on the ASAINR of user 1, i.e., $\gamma_1$. In addition, we set $P_1=P_2=10$, $\sigma^2=1$, $\alpha^2_{1,1}=4$, $\alpha^2_{2,1}=2$, and $q^2_{2,1,1}=3$. Then, we have $\gamma_{1,0}=\frac{P_1L\alpha^2_{1,1}}{\sigma^2+P_2\alpha^2_{2,1}}=40/21=1.9$. In the first example, we set $q^2_{1,1,1}=8$; whereas in the second example, we set $q^2_{1,1,1}=1$. Based on Propositions \ref{wobf} and \ref{wbf}, it can be verified that $\gamma_1$ will monotonically increase with $M$ in the first example, with or without the associated IRS (or passive beamforming). However, this is not true in the second example, as will be shown next.

\begin{figure}[hbtp]
\centering
\subfigure[Without IRS passive beamforming.]{\includegraphics[width=0.49\textwidth]{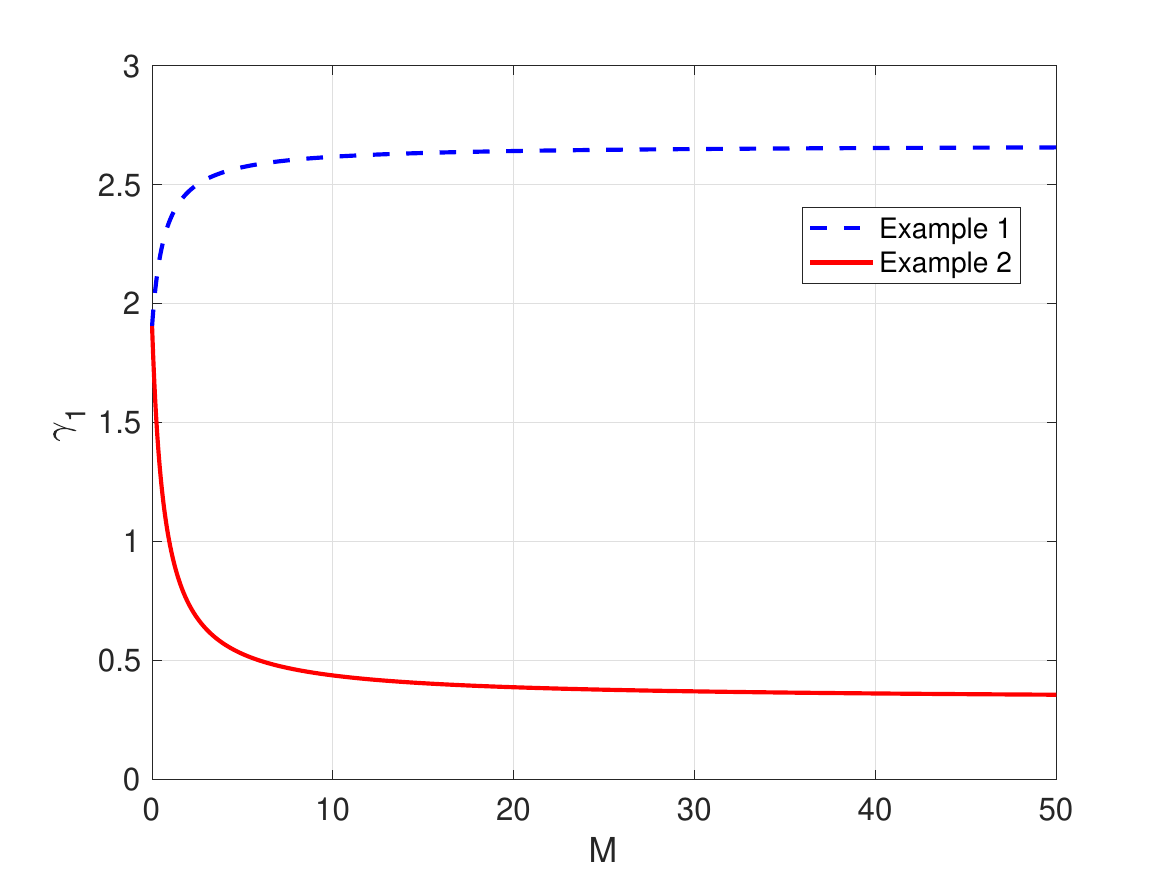}}
\subfigure[With IRS passive beamforming.]{\includegraphics[width=0.49\textwidth]{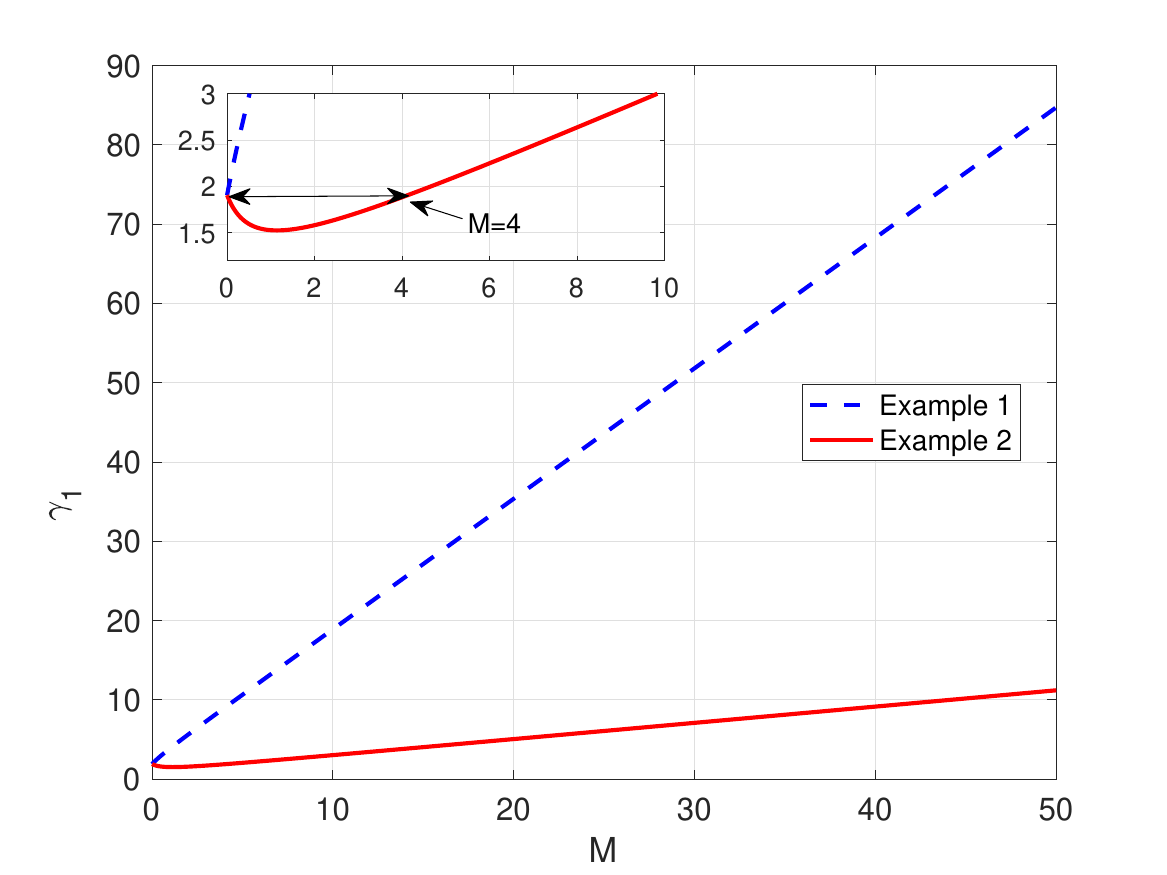}}
\caption{User ASAINR versus number of IRS reflecting elements, $M$.}\label{NumExp}\vspace{-12pt}
\end{figure}
In Fig.\,\ref{NumExp}, we plot $\gamma_1$ versus $M$ for the two above examples, under the two scenarios without versus with IRS passive beamforming. Note that $M=0$ corresponds to the benchmark system without IRS. In the first example, it is observed that $\gamma_1$ monotonically increases with $M$ as expected, without or with IRS passive beamforming. However, in the former scenario, $\gamma_1$ is observed to be upper-bounded by the limit $q^2_{1,1,1}/q^2_{2,1,1}=8/3$ as $M$ increases, in accordance with Proposition \ref{wobf}. In contrast, in the latter scenario, $\gamma_1$ is observed to linearly increase with $M$ even when $M$ is large, in accordance with Proposition \ref{wbf}. In the second example, however, it is observed that without IRS passive beamforming, $\gamma_1$ monotonically decreases with $M$ and is lowered-bounded by the limit $q^2_{1,1,1}/q^2_{2,1,1}=1/3$. Whereas with IRS passive beamforming, $\gamma_1$ is observed to first monotonically decrease with $M$ and then linearly increase with $M$, which is consistent with Proposition \ref{wbf}. In particular, $\gamma_1$ becomes greater than $\gamma_{1,0}$ in the bechmark system without IRS as $M \ge 4$. These two examples show that IRS-user associations have significant effects on the users' ASAINRs and need to be properly designed to ensure that they are linearly increasing with $M$.

\subsection{Network Common ASAINR versus M}\label{pa2}
Next, we consider the maximum network common ASAINR without and with BS power control, i.e., $\gamma_{c,1}^*$ and $\gamma_{c,2}^*$ in (P1) and (P2), respectively. Note that if there is no IRS in the network, the network common ASAINR with the fixed BS transmit powers is given by \[\gamma_0=\mathop {\min}\limits_{k \in {\cal K}}\frac{P_{\max}L\alpha^2_{k,k}}{\sigma^2+\sum\limits_{n \in {\cal K}, n \ne k}P_{\max}\alpha^2_{n,k}}.\]Comparing $\gamma^*_{c,1}$ and $\gamma_0$, we obtain the following proposition.
\begin{proposition}\label{thresprop}
	Suppose that the number of IRSs is no smaller than that of users, i.e., $J \ge K$. Then, we have $\gamma^*_{c,1} > \gamma_0$ if
	\begin{equation}\label{thres3}
	M > \frac{16}{\pi^2\lfloor \frac{J}{K} \rfloor^2}\mathop{\max}\limits_{k \in {\cal K}} \frac{L\alpha^2_{k,k}P_{\max}\sum\limits_{n \in {\cal K}, n \ne k}{\sum\limits_{j \in \cal J}q_{n,j,k}^2}}{q^2_{k,\min}\Big(\sigma^2+\sum\limits_{n \in {\cal K}, n \ne k}P_{\max}\alpha^2_{n,k}\Big)},
\end{equation}
with $q^2_{k,\min} \triangleq \mathop{\min}\limits_{i \in {\cal J}} q^2_{k,i,k}$ denoting the minimum channel power of the reflected BS $k$-user $k$ link via a single IRS in $\cal J$.
\end{proposition}
\begin{IEEEproof}
	Since $J \ge K$, there must exist an IRS-user association strategy, such that each user in $\cal K$ is associated with at least $\lfloor \frac{J}{K} \rfloor$ IRSs in $\cal J$. Consider any one of such strategies and let $\hat\gamma_c$ be the corresponding network common ASAINR. Obviously, we have $\gamma^*_{c,1} \ge \hat\gamma_c$. According to (\ref{thres2g}), if $M$ satisfies
	\begin{equation}\label{thres3g}
		M > \frac{16}{\pi^2}\mathop{\max}\limits_{k \in {\cal K}}\frac{L\alpha^2_{k,k}P_{\max}\sum\limits_{n \in {\cal K}, n \ne k}{\sum\limits_{j \in \cal J}q_{n,j,k}^2}}{\Big(\sum\limits_{j \in \cal J}\lambda_{j,k}q_{k,j,k}\Big)^2\Big(\sigma^2+\sum\limits_{n \in {\cal K}, n \ne k}P_{\max}\alpha^2_{n,k}\Big)},
    \end{equation}
    the individual ASAINR of each user $k, k \in \cal K$ will be greater than that without IRS, so does the network common ASAINR. Thus, we have $\gamma^*_{c,1} \ge \hat\gamma_c > \gamma_0$. Since \[\Big(\sum\limits_{j \in \cal J}\lambda_{j,k}q_{k,j,k}\Big)^2 \ge \Big\lfloor \frac{J}{K} \Big\rfloor^2 q^2_{k,\min}, \forall k \in \cal K,\] it is obvious that (\ref{thres3g}) holds if (\ref{thres3}) is met. The proof is thus completed.
\end{IEEEproof}

In fact, the proof of Proposition \ref{thresprop} implies that as long as (\ref{thres3}) is satisfied and each user in $\cal K$ is associated with at least $\lfloor \frac{J}{K} \rfloor$ IRSs in $\cal J$ (not necessarily optimal), the corresponding network common ASAINR must be larger than $\gamma_0$. However, the above result may not hold if $J<K$. In this case, there always exists at least one user, say user $k, k \in \cal K$, with which no IRS is associated, i.e., $\lambda_{j,k}=0, \forall j \in \cal J$. It follows that if (\ref{mono1}) is satisfied, its receive ASAINR will be degraded  by increasing $M$ compared to the benchmark system without IRS. As a result, there may not exist a universal upper bound on $M$ as in (\ref{thres3}), such that $\gamma^*_{c,1} > \gamma_0$ is ensured.

On the other hand, with BS power control, the optimal BS transmit powers in the no-IRS benchmark system can be derived based on (\ref{pwctrl}) with setting $M=0$. Let $P_{k,0}, k \in \cal K$ represent the optimal transmit power of BS $k$ when $M=0$. By replacing $P_k = P_{\max}$ in (\ref{thres3}) with $P_k = P_{k,0}, k \in \cal K$, we can similarly derive an upper bound on $M$, over which $\gamma^*_{c,2}$ is greater than the maximum network common ASAINR of the no-IRS benchmark system. The details are omitted for brevity.

\section{Proposed Solutions to (P1)}\label{p1}
In this section, we first show that (P1) can be optimally solved by applying the BB algorithm by reformulating it into an MILP problem. In addition, some essential insights into the IRS-user associations are revealed, based on which a successive refinement algorithm of polynomial complexity is also proposed to solve (P1) more efficiently.

\subsection{Optimal Solution by BB Algorithm}\label{bnb}
First, by substituting (\ref{effch1}) and (\ref{mean3}) into the objective function of (P1), the users' ASAINRs can be rewritten as
\begin{align}
\gamma_k({\mv \lambda}_k)&=\frac{\tilde\alpha^2_{k,k}({\mv \lambda}_k)}{\zeta_k}\nonumber\\
&=X_k+\sum\limits_{j \in \cal J}\lambda_{j,k}Y_{j,k}+Z_k\Big(\sum\limits_{j \in \cal J}\lambda_{j,k}q_{k,j,k}\Big)^2, k \in {\cal K},
\end{align}
where we have defined
\begin{align}
&X_k=\frac{1}{\zeta_k}\Big(L\alpha^2_{k,k}+M\sum\limits_{j \in \cal J}{q_{k,j,k}^2}\Big), \;Y_{j,k}  = \frac{MA_{j,k}}{\zeta_k},\nonumber\\
&Z_{k} =\frac{M^2\pi^2}{16\zeta_k}, k \in {\cal K}, j \in {\cal J}.
\end{align}

Then, by introducing a slack variable $z$, (P1) can be reformulated into an equivalent epigraph form, i.e.,
\begin{subequations}\label{op3}
\begin{align}
\mathop {\max}\limits_{z,\mv\Lambda}&\; z\nonumber\\
\text{s.t.}\;\;&X_k+\sum\limits_{j \in \cal J}\lambda_{j,k}Y_{j,k}+Z_k\Big(\sum\limits_{j \in \cal J}\lambda_{j,k}q_{k,j,k}\Big)^2 \ge z, \forall k \in {\cal K},\label{op3a}\\
&\sum\limits_{k \in {\cal K}}{\lambda_{j,k}} \le 1, \forall j \in {\cal J}, \label{op3b}\\
&\lambda_{j,k} \in \{0,1\}, \forall j \in {\cal J}, k \in {\cal K}.\label{op3d}
\end{align}
\end{subequations}

It is worth noting that at the optimality of (\ref{op3}), the minimum of the left-hand side of (\ref{op3a}) must be equal to $z$, since otherwise, we can increase $z$ until they are equal, while achieving a larger objective value of (\ref{op3}). Thus, the optimal $z$ in (\ref{op3}) should be equal to the maximum network common ASAINR in (P1). However, problem (\ref{op3}) remains a non-convex MINLP. As a result, the general BB algorithm, which is only applicable to convex MINLP, cannot be directly applied to solve (\ref{op3}). In view of this challenge, we next reformulate problem (\ref{op3}) into an equivalent convex MILP, by introducing additional auxiliary variables thanks to the unique binary property of $\lambda_{j,k}$'s.

Specifically, for the square term $(\sum\nolimits_{j \in \cal J}\lambda_{j,k}q_{k,j,k})^2$ in (\ref{op3a}), it must hold that
\small\begin{align}
\Big(\sum\limits_{j \in \cal J}\lambda_{j,k}q_{k,j,k}\Big)^2&\!\!=\!\sum\limits_{j \in \cal J}\lambda^{2}_{j,k}q^2_{k,j,k}\!+\!\sum\limits_{i \in \cal J}\sum\limits_{j \in {\cal J}, j \ne i}\!\!\!\lambda_{i,k}\lambda_{j,k}q_{k,i,k}q_{k,j,k}\nonumber\\
&\!\!=\!\sum\limits_{j \in \cal J}\lambda_{j,k}q^2_{k,j,k}\!+\!2\!\sum\limits_{i \in \cal J}\sum\limits_{j \in {\cal J}, j > i}\!\!\!\!\lambda_{i,k}\lambda_{j,k}q_{k,i,k}q_{k,j,k},\label{expansion}
\end{align}\normalsize
where the equality (\ref{expansion}) is due to the fact that $\lambda^2_{j,k}=\lambda_{j,k}, \forall j,k$. To address the second nonlinear term in (\ref{expansion}), we introduce the following auxiliary variables $\psi_{i,j,k}=\lambda_{i,k}\lambda_{j,k}, \forall i,j \in {\cal J}, i < j, k \in {\cal K}$. As such, (\ref{expansion}) becomes affine in $\lambda_{j,k}$'s and $\psi_{i,j,k}$'s.

However, additional $\frac{KJ(J-1)}{2}$ non-convex equality constraints, i.e.,
\begin{equation}\label{neweq}
\psi_{i,j,k}=\lambda_{i,k}\lambda_{j,k}, \forall i,j \in {\cal J}, i < j, k \in {\cal K},
\end{equation}
should be added to problem (\ref{op3}). Fortunately, by exploiting the fact that all $\lambda_{j,k}$'s are binary variables, it can be easily verified that (\ref{neweq}) is equivalent to the following linear inequality constraints:
\begin{equation}\label{trick}
\small\begin{split}
&0 \le \psi_{i,j,k} \le \lambda_{i,k}, \forall i,j \in {\cal J}, j > i, k \in {\cal K},\\
&\lambda_{j,k}+\lambda_{i,k}-1 \le \psi_{i,j,k} \le \lambda_{j,k}-\lambda_{i,k}+1, \forall i,j \in {\cal J}, j > i, k \in {\cal K}.
\end{split}
\end{equation}\normalsize
For example, when $\lambda_{i,k}=1$ and $\lambda_{j,k}=0$, we have $\psi _{i,j,k}=1 \times 0=0$. On the other hand, the first and the second rows of (\ref{trick}) become $0 \le \psi_{i,j,k} \le 1$ and $0 \le \psi_{i,j,k} \le 0$, respectively, which also lead to $\psi_{i,j,k}=0$.

Let $S_k \triangleq \sum\nolimits_{i \in \cal J}\sum\nolimits_{j \in {\cal J}, j > i}\psi_{i,j,k}q_{k,i,k}q_{k,j,k}, k \in \cal K$. By substituting $S_k, k \in {\cal K}$ into (\ref{op3a}) and adding (\ref{trick}) to (\ref{op3}), problem (\ref{op3}) can be equivalently transformed into the following problem,
\begin{subequations}\label{op4}
\begin{align}
\mathop {\max}\limits_{z,{\mv\Lambda},{\mv\Psi}}&\; z \nonumber\\
\text{s.t.}\;\;&X_{k} +\sum\limits_{j \in \cal J}\lambda_{j,k}(Y_{j,k}+Z_{k}q^2_{k,j,k})+2Z_kS_k \ge z, \forall k \in {\cal K},\label{op4a}\\
&\sum\limits_{k \in \cal K}{\lambda_{j,k}} \le 1, \forall j \in {\cal J},\label{op4b}\\
&0 \le \psi_{i,j,k} \le \lambda_{i,k}, \forall i,j \in {\cal J}, j > i, k \in {\cal K},\label{op4c}\\
&{\text{(\ref{trick})}}, \;\lambda_{j,k} \in \{0,1\}, \forall j \in {\cal J}, k \in {\cal K},\label{op4d}
\end{align}
\end{subequations}
where $\mv\Psi=[\psi_{i,j,k}]$ is the ensemble of the auxiliary variables added to problem (\ref{op3}).

Problem (\ref{op4}) is now an MILP which contains $KJ$ binary variables and $\frac{KJ(J-1)}{2}+1$ auxiliary non-binary variables. Thus, this problem can be optimally solved via the BB algorithm, which involves solving a sequence of linear programming problems. Generally, it is difficult to analyze the complexity of the BB algorithm, while its worst-case complexity is equal to that of full enumeration, i.e., ${\cal O}(K^J)$\cite{boyd2007branch}. We show via simulation in Section \ref{sim} that the running time of the former is much less than that of the latter in general.

\subsection{Low-Complexity Solution by Successive Refinement}
Although the optimal solution to problem (\ref{op3}) can be obtained by the BB algorithm, its worst-case complexity, albeit rarely encountered, can still be high with increasing $J$ and/or $K$. To address this issue, in this subsection, we propose a more efficient successive refinement algorithm for (P1) based on a new result pertaining to the effect of IRS-user associations on user ASAINRs. First, we present the following proposition.
\begin{proposition}\label{IRSbalancing}
 	Given any feasible IRS-user association solution $\mv\Lambda$, if an IRS is assigned from user $k'$ to another user $k$ with $k,k' \in \cal K$, then the ASAINRs of users $k$ and $k'$ will increase and decrease, respectively.
\end{proposition}
\begin{IEEEproof}
Please refer to Appendix \ref{appendixB}.
\end{IEEEproof}

Proposition \ref{IRSbalancing} implies that the user ASAINRs can be adjusted by switching the IRS association from one to another user. This thus offers us an efficient successive refinement algorithm to improve the network common ASAINR iteratively. Specifically, let $\mv\Lambda(r)=[{\mv\lambda}_k(r)]$ be the updated IRS-user associations in the $r$-th iteration. Accordingly, the ASAINR achievable by each user $k$ is given by $\gamma_k(r)=\frac{\tilde\alpha^2_{k,k}({\mv \lambda}_k(r))}{\zeta_k}, k \in \cal K$. In the $(r+1)$-th iteration, we first identify the user that achieves the lowest ASAINR among all users given $\mv\Lambda=\mv\Lambda(r)$, referred to as the bottleneck user and denoted as $k_b = \arg \mathop {\min}\limits_{k \in \cal K} \gamma_k(r)$. Obviously, if all IRSs have been assigned to user $k_b$, then its ASAINR cannot be further improved. As such, the optimal IRS-user associations should assign all IRSs in $\cal J$ to user $k_b$, and the iteration can be terminated. This may happen when, for example, the bottleneck user is severely blocked and the number of IRSs is small. Otherwise, according to Proposition \ref{IRSbalancing}, its achieved ASAINR can be improved by assigning it with one more IRS from another user, who, however, will achieve a lower ASAINR. If the assigned IRS is properly chosen to control the increase and decrease in the ASAINRs of these two users, respectively, the network common ASAINR could be improved. Mathematically, denote by $\Omega(r)=\{j | \lambda_{j,k}(r)=1, k \in {\cal K}, k \ne k_b\} \ne \emptyset$ the set of IRSs assigned to the other $K-1$ users after the $r$-th iteration. If an IRS $j, j \in \Omega(r)$ is assigned from its associated user $k'$ (with $\lambda_{j,k'}(r)=1$ and $k' \ne k_b$) to user $k_b$, let $\mv\Lambda(r,j)=[\mv\lambda_k(r,j)]$ denote the corresponding IRS-user associations. Obviously, we have
\begin{equation}\label{update1}
	\mv\lambda_k(r,j)=
	\begin{cases}
		\mv\lambda_k(r)-{\mv e}_j, &{\text{if}}\;k=k'\\
		\mv\lambda_k(r)+{\mv e}_j, &{\text{if}}\;k=k_b\\
		\mv\lambda_k(r), &{\text{otherwise}},
	\end{cases}k \in {\cal K}.
\end{equation}
Note that except users $k'$ and $k_b$, the ASAINRs of all other users in $\cal K$ are not changed after this change of assignment. This is because their associated and non-associated IRSs remain the same as before the above change of assignment. As such, the network common ASAINR is updated as
\begin{equation}\label{update2}
	\gamma_c(r,j)=\mathop {\min}\Big\{\gamma_{k'}(r,j),\gamma_{k_b}(r,j), \mathop {\min}\limits_{k \in {\cal K}\backslash\{k', k_b\}}\!\gamma_k(r)\Big\},
\end{equation}
where $\gamma_{k'}(r,j)$ and $\gamma_{k_b}(r,j)$ denote the ASAINRs of users $k'$ and $k_b$ after this change of assignment, respectively.

Based on (\ref{update2}), in the $(r+1)$-th iteration, we assign the bottleneck user $k_b$ with an IRS that yields the greatest improvement in the network common ASAINR among all IRSs in $\Omega(r)$, denoted as $j^*(r)=\arg \mathop {\max}\limits_{j \in \Omega(r)}\gamma_c(r,j)$. If $j^*(r)$ is not unique, we can choose the one in $j^*(r)$ that yields the greatest improvement in user $k_b$'s ASAINR, i.e., $j^*(r):=\arg \mathop {\max}\limits_{j \in j^*(r)}\gamma_{k_b}(r,j)$. Then, the IRS-user associations should be updated as $\mv\Lambda(r+1)=[\mv\lambda_k(r,j^*(r))]$. The above process proceeds until the network common ASAINR cannot be improved any more. The main procedures of the above algorithm are summarized below in Algorithm \ref{Alg1}. Since $\gamma_c(r)$ is monotonically non-decreasing with $r$, Algorithm \ref{Alg1} is ensured to converge. Moreover, in each iteration, $\lvert \Omega(r) \rvert \le J$ comparisons should be made. Hence, the worst-case complexity of Algorithm \ref{Alg1} is in the order of ${\cal O}(JS_1)$ with $S_1$ denoting its iteration number, which is thus linear w.r.t. $J$.
\begin{algorithm}
  \caption{Successive Refinement Algorithm for Solving (P1)}\label{Alg1}
  \begin{algorithmic}[1]
    \State Let $r=0$. Initialize $\mv\Lambda$ as $\mv\Lambda(r)$ and compute the resulting network common ASAINR as $\gamma_c(r)$.
    \While {convergence is not reached}
    \State Determine the bottleneck user $k_b$ and the set of IRSs assigned to all other users, $\Omega(r)$.
    \If {$\Omega(r)=\emptyset$}
    \State Stop and output $\mv\Lambda(r)$.
    \EndIf
    \State Compute the network common ASAINRs $\gamma_c(r,j), j \in \Omega(r)$ based on (\ref{update1}) and (\ref{update2}).
    \State Determine the best assigned IRS to user $k_b$ as $j^*(r)=\arg \mathop {\max}\limits_{j \in \Omega(r)}\gamma_c(r,j)$.
    \If {$j^*(r)$ is not unique}
    \State Determine $j^*(r)$ as $j^*(r):=\arg \mathop {\max}\limits_{j \in j^*(r)}\gamma_{k_b}(r,j)$.
    \EndIf
    \State Update the network common ASAINR as $\gamma_c(r+1)=\gamma_c(r,j^*(r))$.
    \If {$\gamma_c(r+1)>\gamma_c(r)$}
    \State Update $\mv\Lambda(r+1)=[\mv\lambda_k(r,j^*(r))]$.
    \Else
    \State Stop and output $\mv\Lambda(r)$.
    \EndIf
    \State Update $r=r+1$.
    \EndWhile
    \end{algorithmic}
\end{algorithm}

\section{Proposed Solution to (P2)}\label{joint}
In this section, we focus on solving the more challenging problem (P2) with joint optimization of IRS-user associations $\mv \Lambda$ and BS transmit powers $\mv P$. To avoid the early-termination issue of the AO as discussed in Section \ref{pf}, we first reformulate (P2) into an equivalent problem with only the IRS-user association variables in $\mv\Lambda$. Specifically, as the optimal value of (P2) with any given $\mv\Lambda$ is available in (\ref{comSINR}), (P2) is equivalent to maximizing (\ref{comSINR}) or minimizing the reciprocal of (\ref{comSINR}) by optimizing $\mv\Lambda$ only, i.e.,
\begin{subequations}\label{op5}
\begin{align}
\mathop {\min}\limits_{\mv\Lambda}\quad &\mathop {\max}_{k \in \cal K}\;\rho\left(\tilde{\mv F}(\mv\Lambda)+\frac{1}{P_{\max}}\tilde{\mv v}(\mv\Lambda){\mv e}^T_k\right)\nonumber\\
\text{s.t.}\;\;&\sum\limits_{k \in {\cal K}}{\lambda_{j,k}} \le 1, \forall j \in {\cal J}, \\
&\lambda_{j,k} \in \{0,1\}, \forall j \in {\cal J}, k \in {\cal K}.
\end{align}
\end{subequations}
Compared to (P2), only $\mv\Lambda$ needs to be optimized in problem (\ref{op5}). Nonetheless, it is generally difficult to express its objective function in an analytical form of $\mv\Lambda$, thus making problem (\ref{op5}) intractable to solve. Besides, the proposed successive refinement algorithm in Algorithm 1 cannot be used to solve (\ref{op5}) either. This is because with BS power control, all users in $\cal K$ always achieve the same ASAINR, as given in (\ref{comSINR}). As a result, assigning one or more IRSs from one user to another user will affect all users' ASAINRs. This is fundamentally different from the case without BS power control, where only the two pertinent users' ASAINRs will be affected.

To handle this difficulty, we propose a sequential update algorithm to solve problem (\ref{op5}) iteratively, by sequentially updating the associated user of each IRS $j, j \in \cal J$ (i.e., the $j$-th row of $\mv\Lambda$) while fixing those of all other IRSs until no performance improvement can be achieved. Note that the early-termination issue with AO is resulted as $\mv P$ and $\mv\Lambda$ are coupled in the iteration. However, the proposed sequential update algorithm circumvents the effect of $\mv P$, thus avoiding the early-termination issue. Specifically, denote by ${\mv \Lambda}_j = [\lambda_{j,1},\lambda_{j,2},\cdots,\lambda_{j,K}], j \in \cal J$ the $j$-th row of ${\mv \Lambda}$. We aim to sequentially optimize ${\mv\Lambda}_j, j \in \cal J$ with the other $J-1$ rows of ${\mv \Lambda}$ being fixed, i.e.,
\begin{align}
\mathop {\min}\limits_{{\mv \Lambda}_j}\;& \mathop {\max}_{k \in \cal K}\;\rho\left(\tilde{\mv F}({\mv \Lambda}_j;\{{\mv \Lambda}_i\}_{i \ne j})+\frac{1}{P_{\max}}\tilde{\mv v}({\mv \Lambda}_j;\{{\mv \Lambda}_i\}_{i \ne j}){\mv e}^T_k\right)\nonumber\\
\text{s.t.}\;\;&\sum\limits_{k \in {\cal K}}\lambda_{j,k} \le 1, \lambda_{j,k} \in \{0,1\}, \forall k \in {\cal K}.\label{sr}
\end{align}

The optimal solution to (\ref{sr}) can be efficiently derived by enumerating the $K$ feasible solutions or possible associated users\footnote{According to Proposition \ref{IRSbalancing}, assigning any IRS to a user can always improve its channel power with its serving BS. As such, there is no need to consider the case with $\lambda_{j,k}=0, \forall k \in \cal K$ in solving (\ref{sr}).} for IRS $j$. To compute the objective value of (\ref{sr}) for each feasible solution, at most $K$ spectral radii need to be computed. Thus, the overall complexity of solving (\ref{sr}) is in the order of ${\cal O}(K^4)$. After solving (\ref{sr}), the associated user of IRS $j$ is updated and the update for the next IRS $j+1$ follows. As the network common ASAINR may be improved by sequentially updating the associated user of each IRS, this process produces a non-decreasing objective value of (\ref{op5}) and thus, the convergence is guaranteed. The update proceeds until the network common ASAINR cannot be further improved by updating any of the IRS in $\cal J$. If each IRS in $\cal J$ is updated $S_2$ times in total, the worst-case complexity of this sequential update algorithm is ${\cal O}(K^4JS_2)$. To further reduce its complexity, notice that $M$ is usually large in practice; hence, it usually holds that $P_{\max}\tilde\alpha^2_{k,k} \gg \sigma^2, \forall k \in \cal K$, leading to $\frac{1}{P_{\max}}\tilde{\mv v}(\mv \Lambda) \rightarrow \bf 0$. As a result, instead of computing $K$ spectral radii in each iteration to solve (\ref{sr}), only a single spectral radius of $\rho(\tilde{\mv F}({\mv \Lambda}_j;\{{\mv \Lambda}_i\}_{i \ne j}))$ needs to be calculated, which reduces the worst-case complexity to ${\cal O}(K^3JS_2)$. This is referred to as the simplified sequential update algorithm in the sequel of this paper. It is shown via simulation in Section \ref{sim} that both the sequential update algorithm and its simplified version are able to achieve near-optimal performance.
\begin{remark}\label{algcomp}
It should be mentioned that the sequential update algorithm can also be applied to solve (P1). Specifically, when IRS $j, j \in \cal J$ needs to be updated, the network common ASAINR by assigning it from the associated user to the bottleneck user is computed and compared with the incumbent. In this regard, the sequential update algorithm incurs a comparable worst-case complexity to the successive refinement algorithm in solving (P1). However, as will be shown in Section \ref{sim}, the former generally yields a worse performance than the latter. The reason is that in the latter algorithm, the bottleneck user is always assigned with the best IRS that achieves the greatest improvement in the network common ASAINR in each iteration. Whereas in the former algorithm, the IRSs assigned to the bottleneck user have to follow the prescribed order of the update for the IRSs. Thus, it is more likely to get stuck in a low-quality suboptimal solution.
\end{remark}

\section{Numerical Results}\label{sim}
In this section, numerical results are provided to validate our performance analysis in Section \ref{pa} and the efficacy of the proposed algorithms compared to some benchmark schemes. Unless otherwise specified, the simulation settings are as follows. We consider a cellular network with $K=4$ BSs, $K=4$ users in a cluster and $J=8$ IRSs, as shown in Fig.\,\ref{topology}. Each BS is equipped with $L=8$ antennas, while each user is equipped with an omnidirectional antenna. For all the BS-user, BS-IRS and IRS-user links involved, their distance-dependent average power gains follow the path-loss model of the urban macro (UMa) scenario in the 3GPP technical specification\cite{3GPP38901}. The bandwidth of the communication link is set to 10 MHz\cite{3GPP38901}. The carrier frequency is 2 GHz, and the noise power spectrum density at the user receiver is $-$164 dBm/Hz. The height of each BS is set to 15 m, while that of each IRS and user is set to 1.5 m. The total transmit power budget of each BS is 40 dBm.  The IRS-user associations in all iterative algorithms (i.e., the successive refinement algorithm for (P1) and the sequential update algorithm and AO algorithm for (P2)) are initialized based on a nearest association rule, i.e., each IRS is associated with the user which is closest to it among all users in $\cal K$.
\begin{figure}[!t]
\centering
\centering
\subfigure[3D plot]{\includegraphics[height=0.22\textwidth]{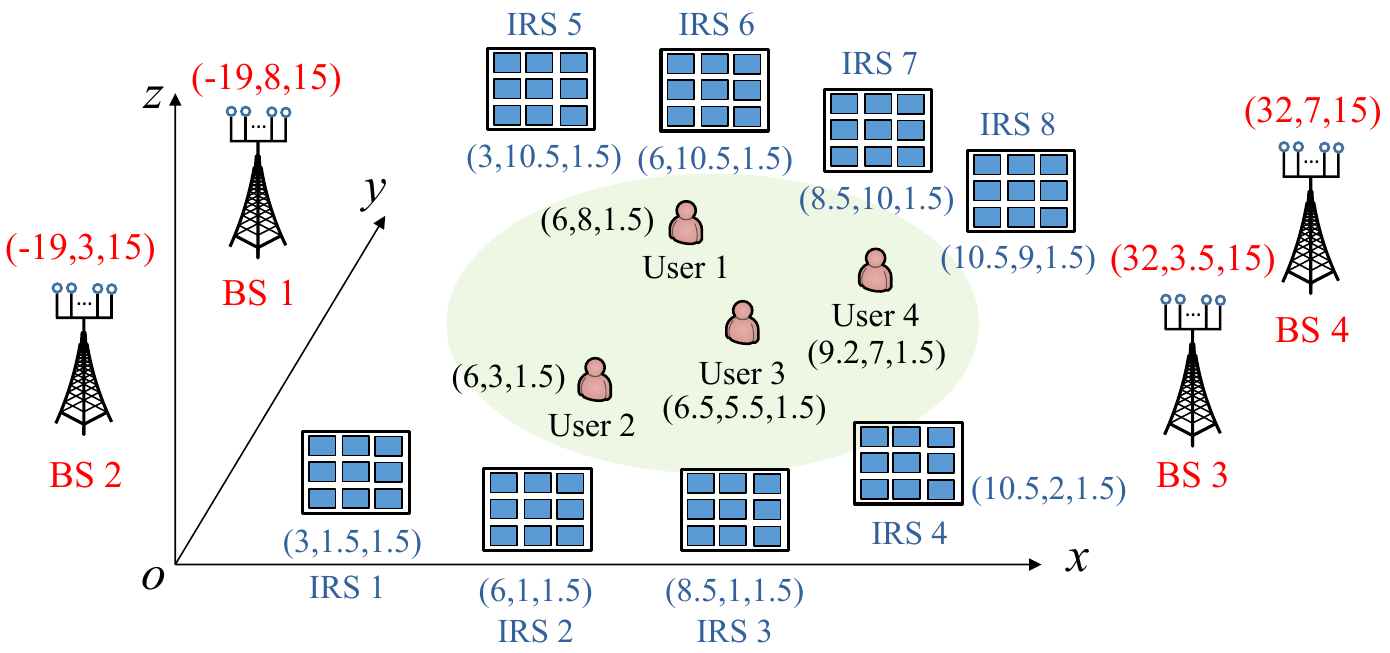}}\;
\subfigure[Top view]{\includegraphics[height=0.22\textwidth]{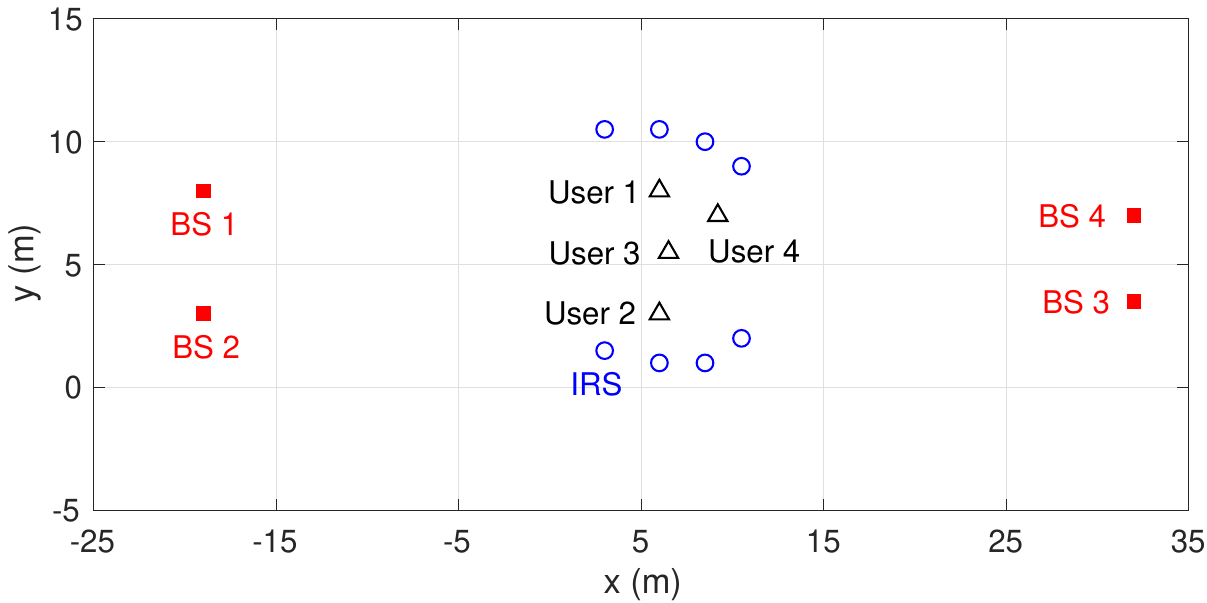}}
\caption{Simulation setup.}\label{topology}
\end{figure}

\begin{table}[htbp]
\centering
\caption{Running time in second of different algorithms for solving (P1)}\label{time}
\begin{tabular}{|l|c|c|c|}\hline
                  & $J=6$ & $J=7$ & $J=8$  \\ \hline
BB algorithm      &  0.18   &   0.39   &   0.57 \\ \hline
Successive refinement      &  0.03   &   0.03   &   0.02 \\ \hline
Full enumeration &  1.07   &   3.58   &  14.33 \\ \hline
\end{tabular}\vspace{-12pt}
\end{table}
First, we evaluate the computational efficiency of the BB algorithm and the successive refinement algorithm as compared to the full enumeration in solving (P1). The number of IRS reflecting elements is set to $M=300$. In Table \ref{time}, we show the running time (in second) of two optimal algorithms for solving (P1) with $J=6, 7$ and $8$. It is observed that the BB algorithm takes much less running time than full enumeration to obtain the optimal solution to (P1), and the saving in time becomes more significant as $J$ increases. In addition, the proposed successive refinement algorithm is observed to take even less running time than the BB algorithm.

\begin{figure}
\centering
\begin{minipage}[t]{0.49\textwidth}
\centering
\includegraphics[width=3.5in]{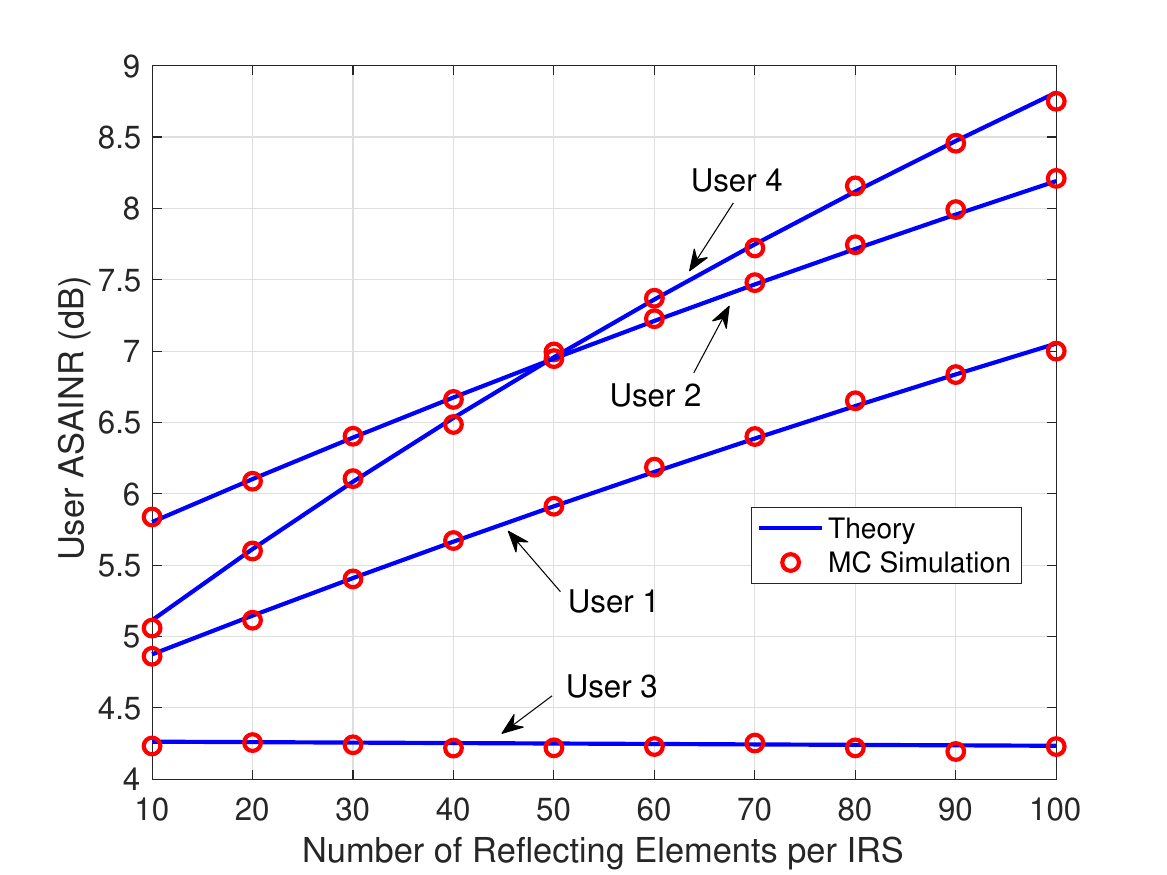}
\caption{Comparison between the theoretical results and the MC simulation results for user ASAINRs.}\label{MCvsTheo}
\end{minipage}
\hfill
\begin{minipage}[t]{0.49\textwidth}
\centering
\includegraphics[width=3.5in]{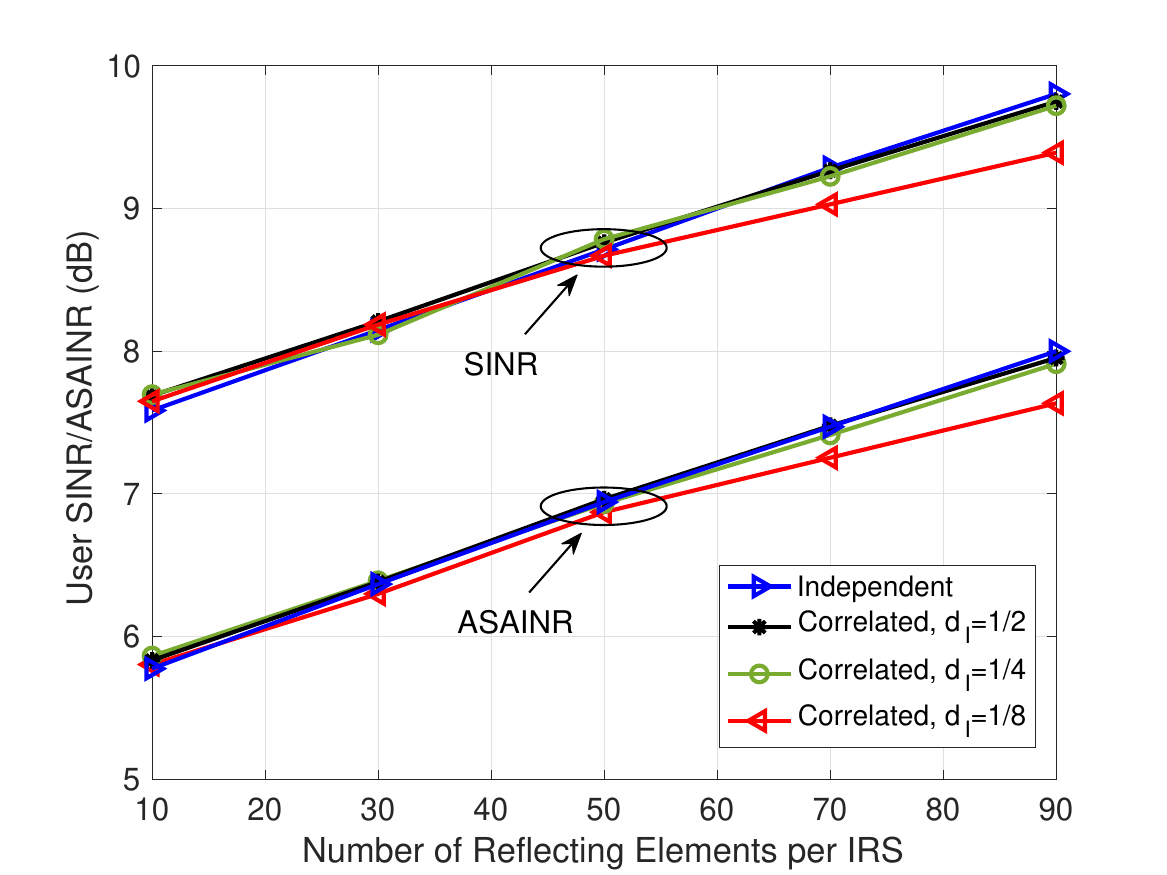}
\caption{Comparison between the results under independent Rayleigh fading and correlated Rayleigh fading}\label{Ele_space}
\end{minipage}
\vspace{-12pt}
\end{figure}
Next, we verify our theoretical results of the user ASAINR in (\ref{mean3}) by Monte-Carlo (MC) simulations. Each MC simulation result is obtained by averaging over 10000 realizations of the BS-user, BS-IRS and IRS-user channels. All BSs are assumed to transmit at their maximum power of 40 dBm, and the IRS-user associations are set based on the nearest association rule. Fig.\,\ref{MCvsTheo} plots the four users' ASAINRs by using (\ref{mean3}) and MC simulations, respectively, under different numbers of IRS reflecting elements $M$. It is observed from Fig.\,\ref{MCvsTheo} that our theoretical results match well with the MC simulation results. This manifests the high approximation accuracy of treating the phase shifts of IRSs as uniformly distributed random variables between $0$ and $2\pi$ for characterizing the IRS random scattering. It is also observed that the ASAINRs of users 1, 2 and 4 monotonically increase with $M$, while that of user 3 slightly decreases with $M$. This is because under the nearest association rule, no IRS can be assigned to user 3, which thus suffers substantial ASAINR loss without reaping the passive beamforming gain. As such, more efficient IRS-user associations are needed to improve its ASAINR.

Under the same setting as in Fig.\,\ref{MCvsTheo}, to evaluate the accuracy of assuming independent Rayleigh fading for all IRS elements, we plot the average SINR and ASAINR of user 2 under practical spatially correlated Rayleigh fading model proposed in \cite{bjornson2020rayleigh} and compare them with those under independent Rayleigh fading in Fig.\,\ref{Ele_space}. Except the ASAINR under independent Rayleigh fading, all results are obtained via the MC simulation. Let $d_I$ denote the spacing between two adjacent IRS elements in both vertical and horizontal directions, which is normalized by the wavelength. It is observed from Fig.\,\ref{Ele_space} that when $d_I=1/2$ or $d_I=1/4$, assuming independent Rayleigh fading achieves a high approximation accuracy for both average SINR and ASAINR over the whole range of $M$ considered. It is also observed that with a given $M$ and $d_I$, the average SINR is larger than the ASAINR, which is consistent with (\ref{sinr.lw}).

\begin{figure}[!t]
\centering
\subfigure[Without BS power control.]{\includegraphics[width=0.45\textwidth]{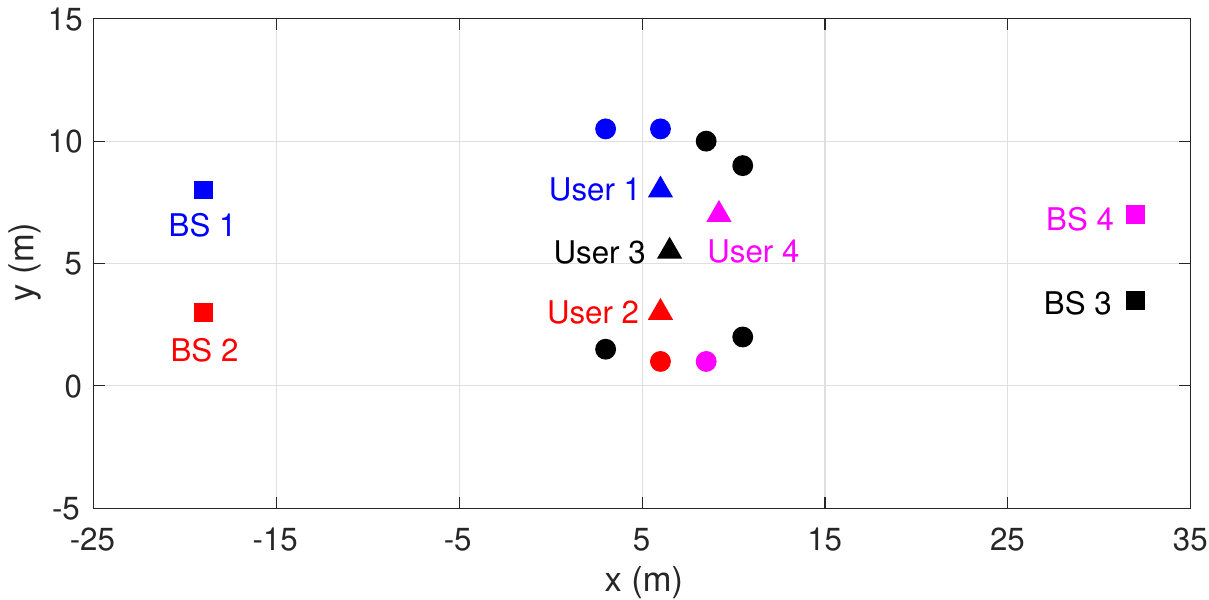}}
\subfigure[With BS power control.]{\includegraphics[width=0.45\textwidth]{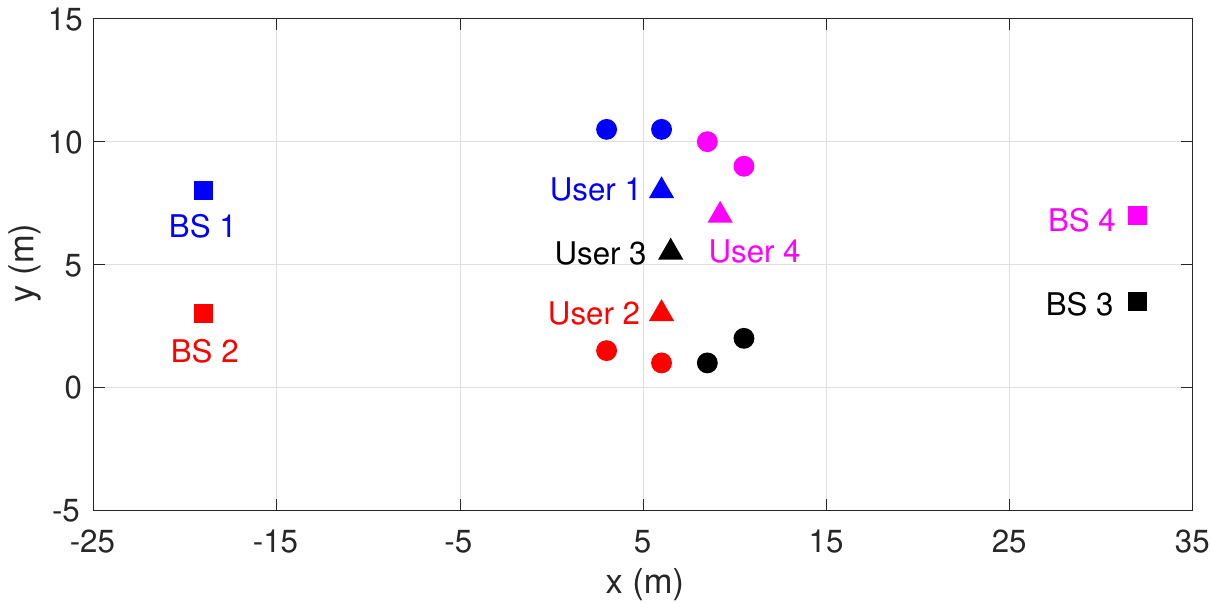}}
\caption{Optimized IRS-user associations with versus without BS power control.}\vspace{-12pt}\label{DLAssocAll}
\end{figure}
Next, we plot the optimized IRS-user associations in Fig.\,\ref{DLAssocAll} with versus without BS power control, by the sequential update algorithm and the BB algorithm, respectively. The number of IRS reflecting elements is set to $M=50$. It is worth mentioning that the proposed successive refinement algorithm yields the same IRS-user association solution as the BB algorithm, and thus the results of the successive refinement algorithm are omitted here for brevity. Each user is marked with the same color as its associated BS and IRSs. Note that in the benchmark system without IRS, user 3 is the bottleneck user due to its relatively far distance with BS 3 but short distances with other interfering BSs. Accordingly, it is observed from Fig.\,\ref{DLAssocAll}(a) that without BS power control, 4 out of 8 IRSs should be assigned to user 3 to compensate for its ASAINR loss and thereby balance the ASAINRs of all users, even though they may be closer to other users. However, with BS power control, it is observed from Fig.\,\ref{DLAssocAll}(b) that the other three users are assigned with more nearby IRSs as compared to Fig.\,\ref{DLAssocAll}(a) and all IRSs are more evenly assigned to the four users. The reason is that the BS power control has provided more flexibility for ASAINR balancing than IRS-user associations only due to its continuous tuning and significant effects on both the information signal and co-channel interference powers for the users. In this case, the IRS-user associations are mainly used to enhance the quality of the direct links between the users and their respective serving BSs. As such, the nearby IRSs of each user should be exploited to achieve this purpose, instead of compensating for the ASAINR loss of the farther bottleneck user (user 3) as in the case without BS power control.

\begin{figure}[!t]
\centering
\includegraphics[width=3.5in]{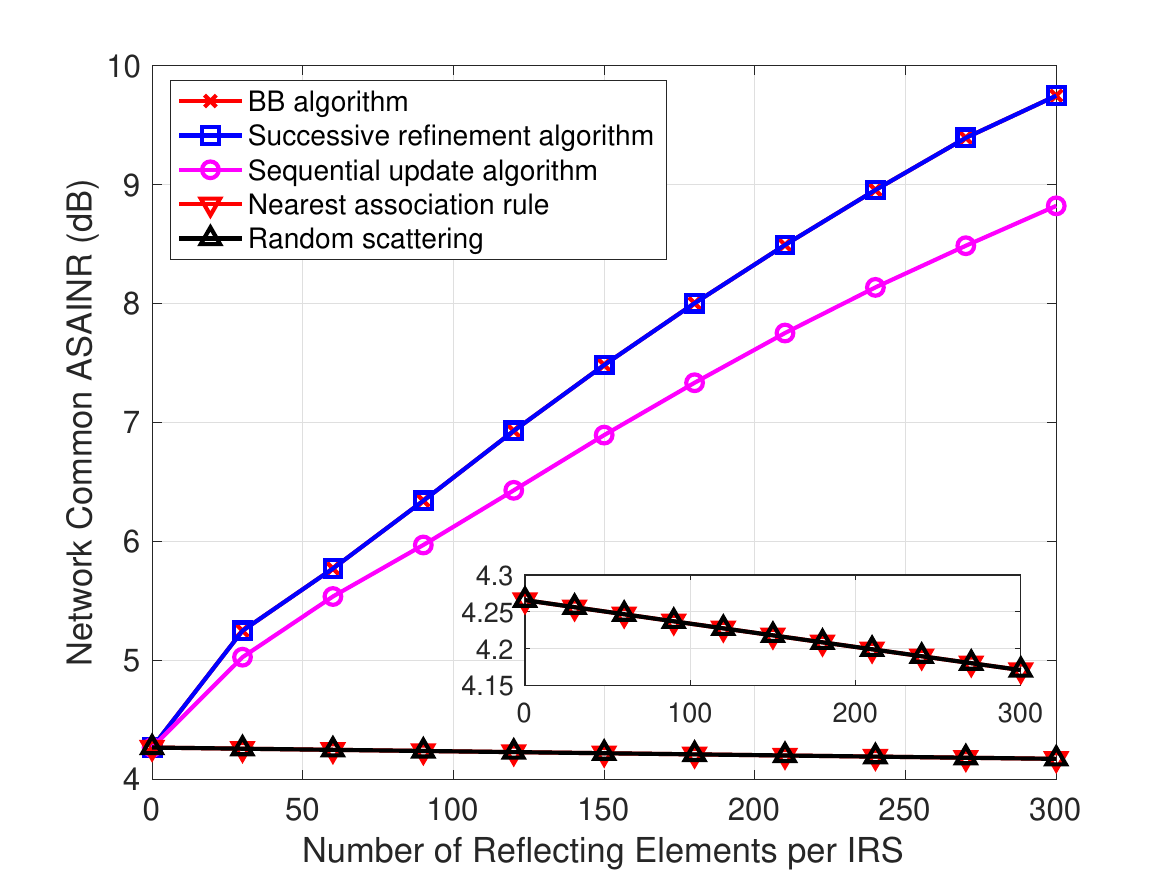}
\DeclareGraphicsExtensions.
\caption{Network common ASAINR without BS power control versus number of IRS reflecting elements.}\label{SINRvsEleNum_npc}
\vspace{-6pt}
\end{figure}
In Fig.\,\ref{SINRvsEleNum_npc}, we plot the network common ASAINRs without BS power control by the proposed BB and successive refinement algorithms as compared to other benchmarks versus the number of IRS reflecting elements, $M$. First, it is observed that the network common ASAINRs by the proposed algorithms monotonically increase with $M$. Moreover, it is observed that the proposed successive refinement algorithm achieves the same performance as the optimal BB algorithm under the considered setup. In accordance with Remark \ref{algcomp}, if the subsequent update algorithm is applied to solve (P1), it achieves a worse performance than the successive refinement algorithm. Furthermore, the nearest association benchmark is observed to only achieve a comparable performance to the IRS random scattering, as no IRS can be assigned to the (bottleneck) user 3 and its ASAINR loss cannot be compensated by IRS passive beamforming. In particular, the achieved network common ASAINRs in these two benchmarks even degrade as $M$ increases. This is consonant with our discussion in Section \ref{pa1}.

\begin{figure}[!t]
\centering
\includegraphics[width=3.5in]{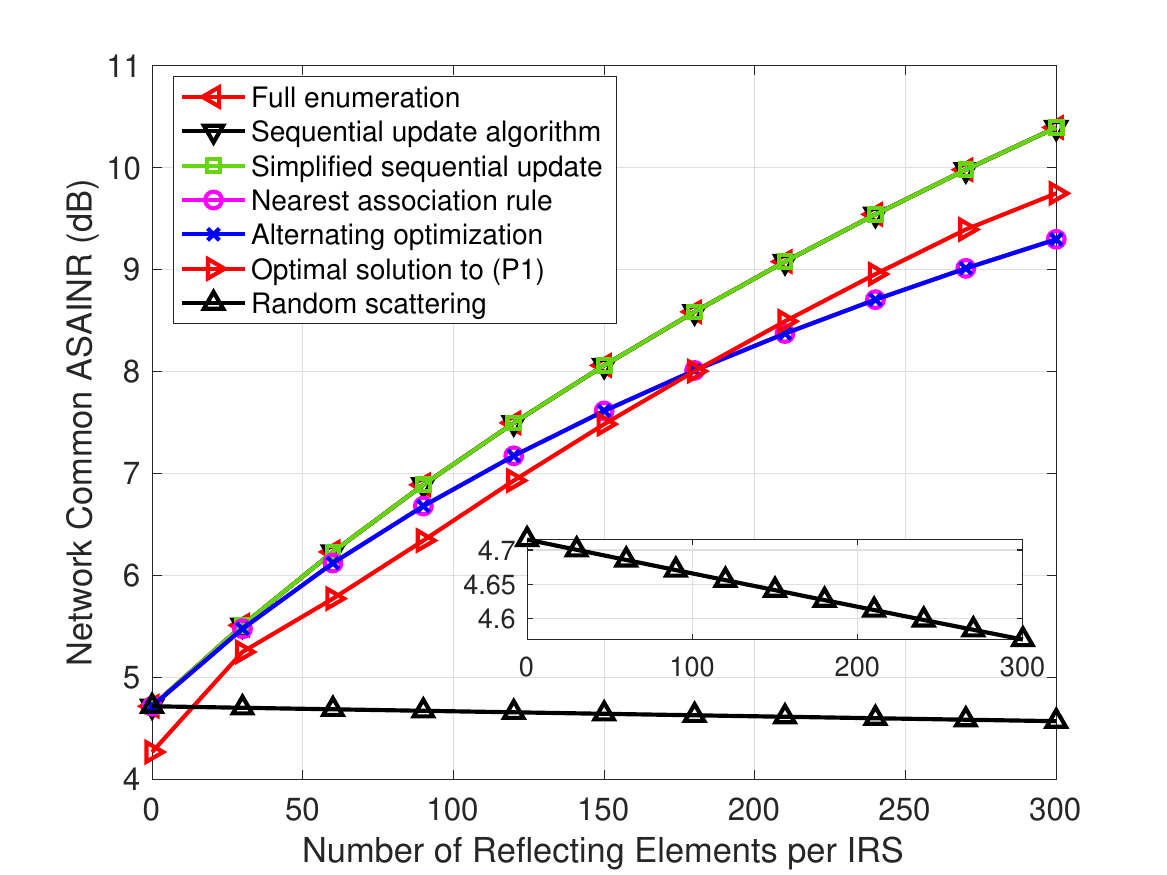}
\DeclareGraphicsExtensions.
\caption{Network common ASAINR with BS power control versus number of IRS reflecting elements.}\label{SINRvsEleNum_pc}
\vspace{-6pt}
\end{figure}
Fig.\,\ref{SINRvsEleNum_pc} shows the network common ASAINRs with BS power control by the proposed sequential update algorithm and its simplified version, as well as other benchmark schemes versus $M$. From Fig.\,\ref{SINRvsEleNum_pc}, it is first observed that both sequential update algorithms proposed can achieve the same performance as the full enumeration and outperform other benchmarks. As such, the simplified sequential update algorithm is more practically appealing due to its lower complexity. Moreover, since no IRS is associated with user 3 in the nearest association benchmark, even with power control, its achieved network common ASAINR is observed to be lower than that without power control (i.e., optimal solution to (P1)) as $M$ becomes large. It is also observed that the AO algorithm only achieves the same performance as the nearest association benchmark, which implies that it fails to update the IRS-user associations beyond the latter due to its early-termination issue shown in Proposition \ref{AOineff}. Finally, with BS power control, it is observed that optimizing IRS-user associations provides much less gain over the nearest association benchmark as compared to Fig.\,\ref{SINRvsEleNum_npc} in the case without BS power control, due to the more flexibility for ASAINR balancing with BS power control.

\begin{figure}[!t]
\centering
\includegraphics[width=3.5in]{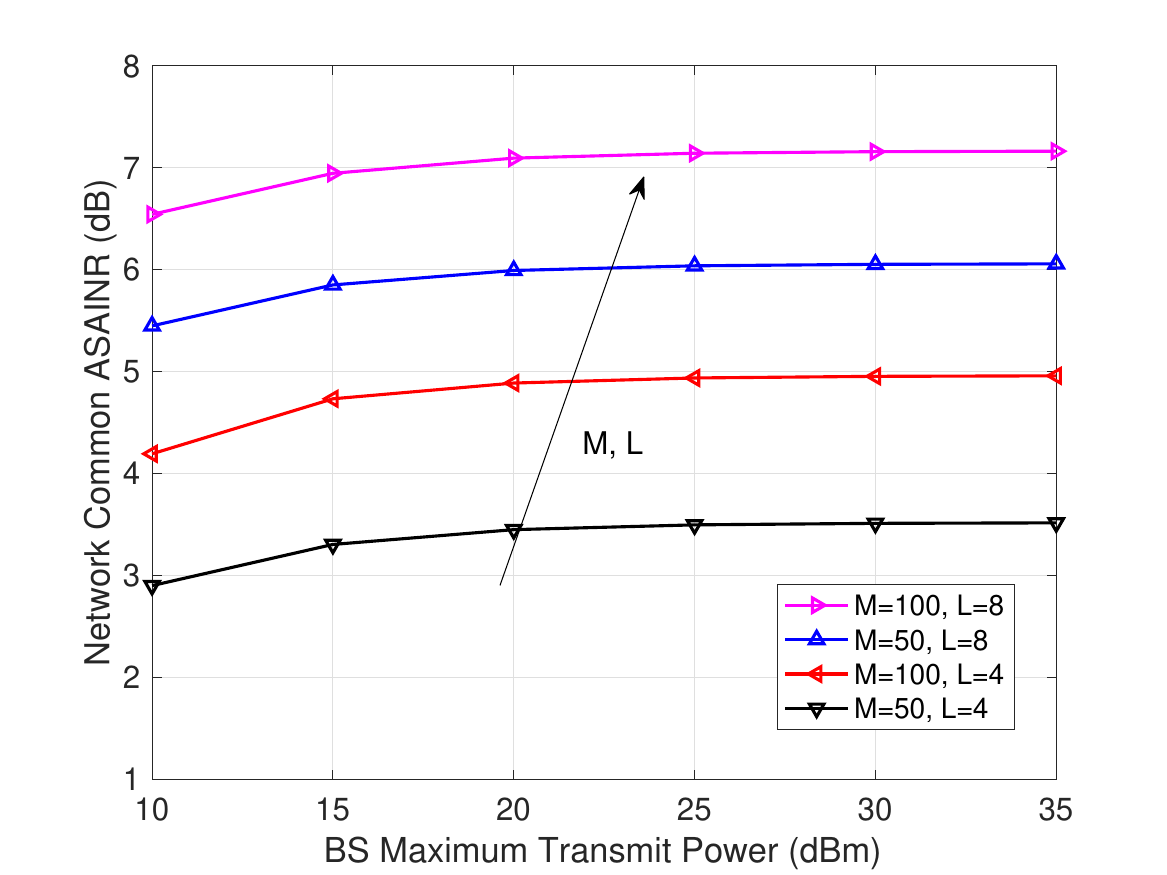}
\DeclareGraphicsExtensions.
\caption{Network common ASAINR versus BS maximum transmit power.}\label{SINRvsPw}
\vspace{-6pt}
\end{figure}
In Fig.\,\ref{SINRvsPw}, we plot the network common ASAINRs with BS power control under different $M$ and $L$ versus BS maximum transmit power $P_{\max}$. It is observed that for any given $M$ and $L$, increasing $P_{\max}$ can hardly improve the network common ASAINR in the high transmit power regime, which is similar to the conventional wireless network without IRS. This is because increasing BS transmit power can potentially enhance the strength of information signal and co-channel interference at each user in a comparable manner, and thus the network common ASAINR will ultimately converge to a limit as $P_{\max}$ becomes large, which is the optimal value of (P2) with $P_{\max} \rightarrow \infty$. On the other hand, it is also observed that for any given $P_{\max}$, increasing $M$ or $L$ can dramatically enhance the network common ASAINR by improving the above ASAINR limit. This is in accordance with our ASAINR analysis in Section \ref{pa} and indicates that increasing $M$ or $L$ is an effective means to enhance the user ASAINRs in the high-$P_{\max}$ or interference-limited regime, while increasing $M$ is generally more energy-efficient and cost-effective due to the passive nature of IRS. Nonetheless, it is shown that increasing $M$ yields less improvement in network common ASAINR under $L=8$ as compared to $L=4$, which is consistent with our discussion below (\ref{thres2}). 

\begin{figure}[!t]
\centering
\includegraphics[width=3.5in]{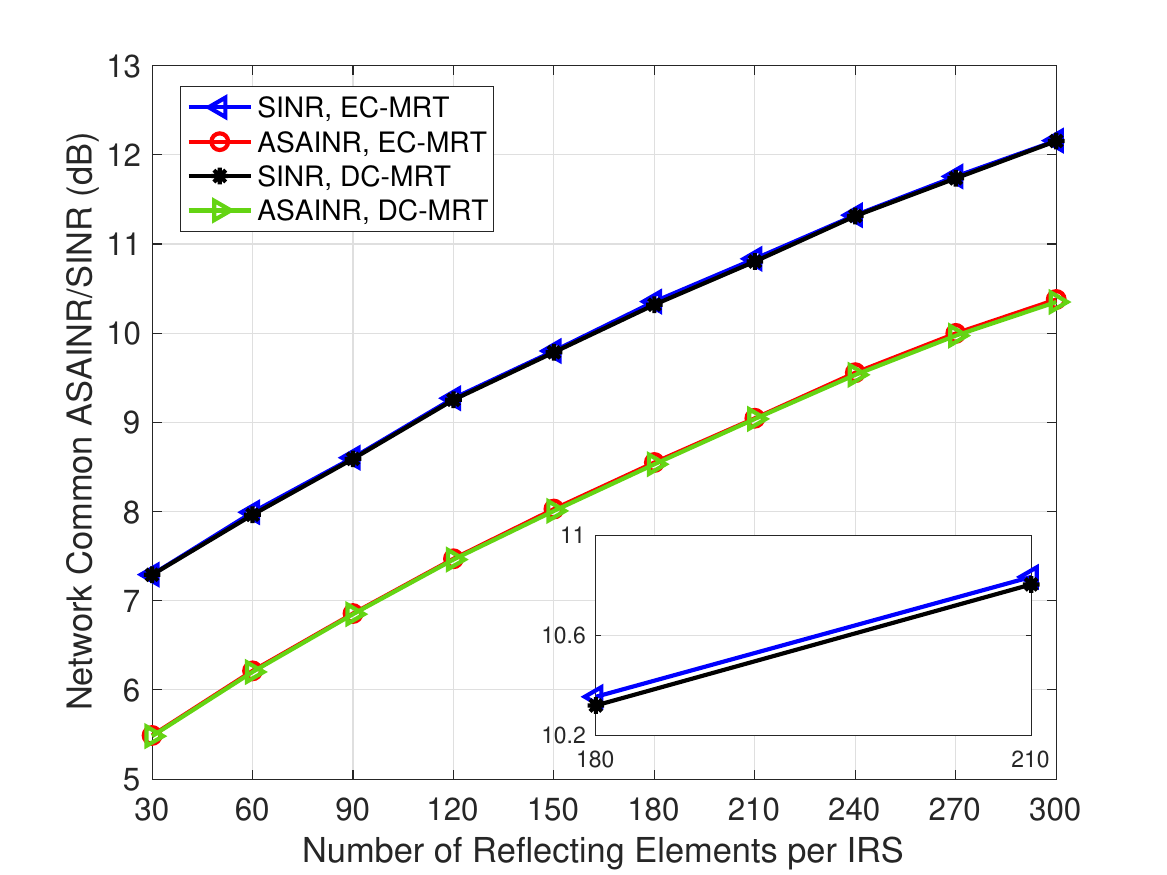}
\DeclareGraphicsExtensions.
\caption{Network common SINR and ASAINR under different MRT designs.}\label{MRT_rd}
\vspace{-6pt}
\end{figure}
Finally, to verify our analysis in Section \ref{metric}, Fig.\,\ref{MRT_rd} shows the network common SINRs and ASAINRs with BS power control under the DC-MRT and EC-MRT (reset after the effective BS-user channels are estimated based on the optimized IRS-user associations). Except the network common ASAINR by the DC-MRT, all results are obtained via the MC simulation. It is observed that the EC-MRT can slightly improve the network common SINR and ASAINR by the DC-MRT, which is in accordance with our analysis following (\ref{sinr.lw}). Nonetheless, the improvement is observed to be marginal, which implies that in the case of Rayleigh-fading BS-IRS channels, using DC-MRT may achieve a near-optimal performance in terms of maximizing the user average SINR/ASAINR. This is in contrast to the case where the BS-IRS channel has a dominant line-of-sight (LoS) component, for which the optimal BS beamforming in general needs to balance between DC-MRT and BS-IRS MRT\cite{wu2019intelligent}. The reason is that in this case, the BS-IRS MRT is able to provide a joint active and passive beamforming gain of $LM^2$ over the cascaded BS-IRS-user link; as a result, the BS-IRS channel plays a more significant role in the optimal active beamforming design at the BS.

\section{Conclusions}
This paper studies two new ASAINR balancing problems in a multi-IRS aided wireless network by optimizing the IRS-user associations with or without the BS power control. The average user ASAINR is first derived in closed form as a lower bound of average user SINR, and it shows different scaling behaviors with the number of IRS reflecting elements, with versus without associated IRSs (or passive beamforming). It is also shown that if the number of IRSs is more than that of users, such that each user can be associated with at least one IRS, there exists a universal bound on the number of IRS reflecting elements, above which the maximum network common ASAINR in the multi-IRS aided wireless network is ensured to be greater than that in the conventional benchmark system without using IRS. Furthermore, an optimal algorithm and two suboptimal algorithms of lower complexity are proposed to solve the formulated problems without the need of performing a full enumeration of IRS-user associations. Numerical results validate our performance analysis and show that both proposed suboptimal algorithms achieve near-optimal performance as compared to the full enumeration. It is also revealed that IRS-user associations have a more significant effect on user ASAINR balancing in the absence of BS power control.

This paper can be extended in several promising directions for future work. First, it is interesting to consider the more general system/channel setup, such as discrete phase shift levels\cite{wu2019discrete} or practical phase shift model\cite{abeywickrama2020intelligent} at each IRS, where the individual user ASAINR and the network common ASAINR are more intricate to be characterized for the optimal IRS-user association design. In addition, some existing works have compared the energy efficiency of the IRS-aided wireless system with that of the conventional relay system\cite{huang2019reconfigurable,bjornson2019intelligent}, yet only in the link level with a single IRS/relay, BS and user. Thus, it is worthy of further comparing their energy efficiency in the network level with multiple IRSs/relays, BSs and users in future work.

\appendix
\subsection{Proof of Proposition \ref{AOineff}}\label{appendixA}
We prove Proposition \ref{AOineff} by contradiction. Suppose that $\mv\Lambda$ can be updated twice and denote by $\mv\Lambda^{(1)}=[{\mv\lambda}^{(1)}_k]$ and $\mv\Lambda^{(2)}=[{\mv\lambda}^{(2)}_k]$ the optimized $\mv\Lambda$ after these two updates, respectively. Moreover, we assume that in the first update, ${\mv\Lambda}$ is updated as $\mv\Lambda^{(1)}$ with $\mv P$ fixed as ${\mv P}_0=[P^{(0)}_k]$. Then, after this update, $\mv P$ is updated as ${\mv P}(\mv\Lambda^{(1)})=[P^{(1)}_k]$ according to (\ref{pwctrl}) and all users achieve the {\it same} ASAINR as \[\gamma^*_{c,2}(\mv\Lambda^{(1)})=\frac{P^{(1)}_k\tilde\alpha^2_{k,k}({\mv \lambda}^{(1)}_k)}{\sigma^2+\sum\limits_{n \in {\cal K}, n \ne k}P^{(1)}_n\nu^2_{n,k}}, k \in \cal K.\]In the second update of $\mv\Lambda$, $\mv\Lambda$ is updated as $\mv\Lambda^{(2)}$ with $\mv P$ fixed as ${\mv P}(\mv\Lambda^{(1)})$. Thus, after the second update, the users' ASAINRs are given by \[\gamma_k(\mv\Lambda^{(2)})=\frac{P^{(1)}_k\tilde\alpha^2_{k,k}({\mv \lambda}^{(2)}_k)}{\sigma^2+\sum\limits_{n \in {\cal K}, n \ne k}P^{(1)}_n\nu^2_{n,k}}, k \in \cal K.\]Obviously, to ensure the successive refinement of the network common ASAINR, it must hold that $\gamma_k(\mv\Lambda^{(2)}) > \gamma^*_{c,2}(\mv\Lambda^{(1)}), \forall k \in \cal K$, or equivalently, $\tilde\alpha^2_{k,k}({\mv \lambda}^{(2)}_k) > \tilde\alpha^2_{k,k}({\mv \lambda}^{(1)}_k), \forall k \in \cal K$. As a result, it also holds that \[\frac{P^{(0)}_k\tilde\alpha^2_{k,k}({\mv \lambda}^{(2)}_k)}{\sigma^2+\sum\limits_{n \in {\cal K}, n \ne k}P^{(0)}_n\nu^2_{n,k}} \ge \frac{P^{(0)}_k\tilde\alpha^2_{k,k}({\mv \lambda}^{(1)}_k)}{\sigma^2+\sum\limits_{n \in {\cal K}, n \ne k}P^{(0)}_n\nu^2_{n,k}}, k \in \cal K.\]This implies that in the first update, ${\mv\Lambda}^{(2)}$ can yield a higher network common ASAINR as compared to ${\mv\Lambda}^{(1)}$. This contradicts the presumption that ${\mv\Lambda}^{(1)}$ is the optimal IRS-user association solution when $\mv P={\mv P}_0$. The proof is thus completed.

\subsection{Proof of Proposition \ref{IRSbalancing}}\label{appendixB}
As the average interference power received by each user $k, k \in \cal K$, i.e., $\sigma^2+P_{\max}\sum\nolimits_{n \in {\cal K}, n \ne k}\nu^2_{n,k}$ is a constant, the IRS-user associations affect its ASAINR only through its channel power gain with BS $k$, $\tilde\alpha^2_{k,k}$. Then, it suffices to prove that $\tilde\alpha^2_{k,k}$ and $\tilde\alpha^2_{k',k'}$ will increase and decrease, respectively, after this assignment. Assume that an IRS $j_0 \in \cal J$ is assigned from user $k'$ to user $k$. Obviously, user $k$/user $k'$ will obtain higher/lower power via passive beamforming by its associated IRSs but lower/higher power via random scattering by its non-associated IRSs. Mathematically, based on (\ref{effch1}), the increase in $\tilde\alpha^2_{k,k}$ is given by
\begin{equation}
\small\begin{split}
\Delta_k&=\frac{M^2\pi^2}{16}q_{k,j_0,k}\left(\sum\limits_{j \in \cal J}\lambda_{j,k}q_{k,j,k}+q_{k,j_0,k}\right)+MA_{j_0,k}\\
&=\frac{M^2\pi^2}{16}q_{k,j_0,k}\left(\sum\limits_{j \in \cal J}\lambda_{j,k}q_{k,j,k}+q_{k,j_0,k}\right)\\
&\qquad\qquad+M\left(\frac{\pi\Gamma(L+\frac{1}{2}) \alpha_{k,k}}{2\Gamma(L)}q_{k,j_0,k}-\frac{\pi^2}{16}q^2_{k,j_0,k}\right)\\
&> \frac{M^2\pi^2}{16}q^2_{k,j_0,k}-\frac{M\pi^2}{16}q^2_{k,j_0,k}=\frac{M\pi^2q^2_{k,j_0,k}}{16}(M-1)\ge 0.\normalsize
\end{split}
\end{equation}
It follows that $\Delta_k$ must be positive, i.e., $\tilde\alpha^2_{k,k}$ is ensured to be improved. Similarly, it can be shown that $\tilde\alpha^2_{k',k'}$ will decrease after this IRS assignment. The proof is thus completed.\vspace{-9pt}

\bibliography{IRSAssoc}
\bibliographystyle{IEEEtran}

\end{document}